\documentclass[12pt]{article}
\usepackage{amsmath}
\usepackage{bbm}
\usepackage{amsfonts}
\usepackage{graphicx,psfrag,epsf}
\usepackage{subcaption}
\usepackage{enumerate}
\usepackage{natbib}
\usepackage{url} 

\usepackage{hyperref}

\usepackage{tikz}
\usepackage{tikz-qtree}
\usetikzlibrary{arrows,decorations.markings}
\usetikzlibrary{shapes.geometric}
\usetikzlibrary{fit,positioning}

\usepackage{fullpage}

\newcommand{\blind}{1}

\usepackage[colorinlistoftodos,textwidth=1.8cm, textsize=tiny]{todonotes}

\newcommand{\wt}{W\&T}

\begin{document}

\def\spacingset#1{\renewcommand{\baselinestretch}%
{#1}\small\normalsize} \spacingset{1}

\def\spacingset#1{\renewcommand{\baselinestretch}%
{#1}\small\normalsize} \spacingset{1}


\if1\blind
{
  \title{\bf Assessing epidemic curves for evidence of superspreading.}
  \author{Joe Meagher\thanks{joe.meagher@ucd.ie}\\
    and \\
    Nial Friel \hspace{.2cm}\\
    Insight Centre for Data Analytics \\
    School of Mathematics and Statistics \\
    University College Dublin}
  \maketitle
} \fi

\if0\blind
{
  \bigskip
  \bigskip
  \bigskip
  \begin{center}
    {\LARGE\bf Assessing epidemic curves for evidence of superspreading.}
\end{center}
  \medskip
} \fi

\bigskip
\begin{abstract}
The expected number of secondary infections arising from each index case, referred to as the reproduction or $R$ number, is a vital summary statistic for understanding and managing epidemic diseases. There are many methods for estimating $R$; however, few explicitly model heterogeneous disease reproduction, which gives rise to superspreading within the population. We propose a parsimonious discrete-time branching process model for epidemic curves that incorporates heterogeneous individual reproduction numbers. Our Bayesian approach to inference illustrates that this heterogeneity results in less certainty on estimates of the time-varying cohort reproduction number $R_t$.
We apply these methods to a COVID-19 epidemic curve for the Republic of Ireland and find support for heterogeneous disease reproduction. Our analysis allows us to estimate the expected proportion of secondary infections attributable to the most infectious proportion of the population. For example, we estimate that the 20\% most infectious index cases account for approximately 75--98\% of the expected secondary infections with 95\% posterior probability.
In addition, we highlight that heterogeneity is a vital consideration when estimating $R_t$.
\end{abstract}

\noindent%
{\it Keywords:} Heterogeneous disease reproduction, Branching processes, Time-varying reproduction numbers.
\vfill

\newpage

\section{Introduction}
\label{sec:intro}

\textit{Superspreading}, where some individuals give rise to large numbers of secondary infections whilst the majority result in very few or none, is a feature of many epidemics \citep{may1987transmission, shen2004superspreading, wong2020evidence}. This phenomenon is a consequence of heterogeneous disease reproduction, whereby the expected number of secondary infections arising from each index case varies from one individual to the next \citep{lloyd2005superspreading}.

Factors that drive heterogeneity separate into two broad categories \citep{becker1999statistical}.
The first affects the infectiousness of each individual, making their infection more or less transmissible. The second is based on the underlying community structure, influencing the number of infectious individuals' contacts. 
Within these categories, there is a myriad of host, pathogen, and environmental factors that contribute to the expected number of secondary infections from each index case \citep{anderson1992infectious}. 
Thus, many studies think of the infectiousness of individuals within a population as distributed along a continuum \citep{lloyd2005superspreading, grassly2008mathematical, britton2010stochastic}.

Heterogeneous disease reproduction is often manifest empirically as a 20/80 rule, whereby the 20\% most infectious index cases are typically responsible for \textit{at least} 80\% of disease transmission \citep{woolhouse1997heterogeneities}. This phenomenon has significant consequences for the design of control measures to curb the spread of infection. Theoretical analyses show that interventions targeting ``core'' groups of the most infectious individuals within a population yield significant reductions in overall transmission \citep{anderson1992infectious, hadeler1995core, woolhouse1997heterogeneities, lloyd2005superspreading, wallinga2010optimizing, britton2020mathematical}. In practice, however, identifying core groups, a priori, is difficult, if not impossible, and so rigorous contact tracing is required to inform decision making during ongoing epidemics \citep{koopman2004modeling, wallinga2010optimizing}.

Mathematical and statistical modelling offering both ``nowcasts'' describing current disease dynamics and ``forecasts' projecting the trajectory of epidemic curves into the future informed public health interventions during the Coronavirus Disease 2019 (COVID-19) pandemic \citep{flaxman2020estimating}. Heterogeneity poses enormous challenges to these efforts, even for simple models based on a constant reproduction number, denoted \(R\) \citep{forsberg2008likelihood, cori2013new, johnson2020disease, donnat2020modeling}. Although the definition of \(R\) as the expected number of secondary infections for each index case is unchanged \citep{fraser2007estimating}, heterogeneity can drastically widen credible intervals associated with nowcasts of \(R\) \citep{johnson2020disease}. Furthermore, heterogeneity will affect the magnitude of both best- and worst-case scenario forecasts \citep{lloyd2005superspreading, donnat2020modeling}. These effects can be significant even before considering incomplete and delayed reporting of disease incidence.

Interestingly, many state-of-the-art methods for estimating time-varying reproduction numbers do not allow for heterogeneous disease reproduction \citep{wallinga2004different, forsberg2008likelihood, cori2013new, thompson2019improved, gostic2020practical, cori2020epiestim, bertozzi2020challenges}.
Recently \cite{donnat2020modeling}, \cite{johnson2020disease} and \cite{schmidt2020inference} have explored the impact of heterogeneity on estimates for \(R\), a concept that extends directly to the instantaneous time-varying reproduction number \citep{fraser2007estimating, cori2013new}.  

In this report, we extend parsimonious branching process models for the spread of infectious disease developed by \cite{wallinga2004different} and \cite{bertozzi2020challenges} to allow for the heterogeneous disease reproduction described by \cite{lloyd2005superspreading}. Furthermore, Bayesian inference for this model offers estimation and uncertainty quantification for \(R_t\) \citep{hoffman2014no, carpenter2017stan}.
These methods allow us to assess epidemic curves for evidence of superspreading and explore the implications of heterogeneous disease reproduction. As a case study, we consider the COVID-19 epidemic in the Republic of Ireland (Ireland) between July and November, 2020. Our analysis allows us to draw conclusions about the expected proportion of secondary infections attributable to the most infectious individuals. For example, we estimate that the 20\% most infectious individuals in this epidemic give rise to 75--98\% of expected secondary infections with 95\% posterior probability, while 62--82\% of individuals are not expected to pass on the infection, again with 95\% posterior probability.

In section \ref{sec:background} we present theory underpinning the current state-of-the-art branching process models for the spread of disease. We then extend these models to heterogeneous disease reproduction in section \ref{sec:methods} before outlining a Bayesian framework for inference and model comparison. This approach allows us to estimate \(R_t\) and explore the impact of heterogeneity on the distribution of secondary infections. Our case study is described in section \ref{sec:data}. In section \ref{sec:results} we present the results of our analysis and we conclude in section \ref{sec:discussion}.

\section{Background}
\label{sec:background}

Branching process epidemic models offer a flexible set of tools for the analysis of epidemics. Unlike compartmental epidemic models, which separate a population into groups depending on their disease and risk status \citep{kermack1927contribution, keeling2011modeling, bjornstad2020modeling}, these stochastic processes model individual infections within a population \citep{keeling2005networks}. 
Such epidemic models offer robust, parsimonious approaches to estimating time-varying reproduction numbers and quantifying heterogeneity in the reproduction of disease from one individual to the next \citep{wallinga2004different, lloyd2005superspreading, bertozzi2020challenges}. 

Here, we define the time-varying reproduction number \(R_t\), referred to as the \textit{cohort reproduction number}, as the expected number of secondary infections arising from each index case within the cohort of infections recorded at time \(t\). The critical value for the reproduction number lies at \(R_t = 1\). The epidemic grows exponentially when \(R_t > 1\) while sustaining \(R_t < 1\) for a sufficiently long period ensures that the epidemic will eventually die out, provided no new infections come from outside the population. 
\(R_t\), also known as the case reproduction number, is distinct from the \textit{instantaneous reproduction number, which we denote $R_t^*$}, although \(R_t^*\) has the same critical value at \(1\). \(R_t^*\) is the expected number of secondary infections arising from an index case at time \(t\) should conditions remain unchanged for the duration of their infection \citep{fraser2007estimating}. This quantity can be estimated in real-time, as described by \cite{cori2013new, thompson2019improved}, and \cite{johnson2020disease}, as it is a function of infections occurring up to time \(t\). Reliable estimates for \(R_t\), on the other hand, depend on infections occurring after time \(t\) and so must be computed retrospectively. This is discussed in detail by \cite{gostic2020practical}.

When modelling the spread of disease within a population as a continuous-time branching process, consider the epidemic \(\boldsymbol t = \left( t_0, t_1, \dots t_N \right)^\top\), where \(t_0 = 0\) is the index case seeding the epidemic and \(t_i \in \left(0, T \right]\) is the time at which the \(i^{th}\) infection is recorded such that \(t_i < t_{i+1}\) for \(i = 1, \dots, N\). 
The branching process allows for two types of index case. The first type contracts their infection from outside the population of interest and is said to have been imported. For simplicity, we let imported infections follow a Poisson process with a constant intensity \(\mu > 0\). All remaining index cases are secondary infections that arise locally within the population. In the general case, we model secondary infections from index case \(i\) as a Poisson process with intensity \(\beta \left(t, \tau_i \mid \theta_i \right) \geq 0\), a function of calendar time \(t\) and time since infection \(t - t_i = \tau_i > 0\) parameterised by \(\theta_i \in \Theta\). Here, \(t_i\) corresponds to the time case \(i\) is infected and so \(\beta \left(t, \tau_i \mid \theta_i \right)\) models the generation interval between infector-infectee pairs while \(\theta_i\) is a set of marks for case \(i\) allowing the intensity function adapt to environmental changes or individual characteristics. For this branching process, the conditional intensity at time \(t\) is
\begin{equation}
    \lambda^* \left( t \right) = \mu + \sum_{t_j < t} \beta \left(t, \tau_j \mid \theta_j \right).
    \label{eq:conditional_intensity}
\end{equation}

Expectation-Maximisation offers a well established approach to maximum likelihood estimation for branching process models \citep{dempster1977maximum, veen2008estimation, bertozzi2020challenges}. These methods rely on the fact that for any branching process epidemic model there exists a unique transmission network linking infector-infectee pairs such that
\begin{equation}
    p_{ji} = \frac{\beta \left(t_i, t_i - t_j \mid \theta_j \right)}{\lambda^* \left( t_i \right)},
    \label{eq:p_ji}
\end{equation}
is the relative likelihood that \(i\) is a secondary infection of \(j\) for all \(j = 0, \dots, i-1\) and
\begin{equation}
    p_{ii} = \frac{\mu}{\lambda^* \left( t_i \right)},
    \label{eq:p_ii}
\end{equation}
is the relative likelihood that \(i\) was imported (see Appendix \ref{app:branching} for more detail).
This insight allows the relative likelihoods defined by equations (\ref{eq:p_ji}) and (\ref{eq:p_ii}) to serve as a basis for estimating \(R_t\) by counting the expected number of secondary infections associated with each index case.

The approach to this problem adopted by \cite{wallinga2004different} (henceforth referred to as \wt{}) is to first assume that imported cases have been identified a priori. This means that the contribution of \(\mu\) in equation (\ref{eq:p_ji}) can be ignored and \(p_{ii} = 0\) for all remaining index cases. A further assumption made by \wt{} is that \(\beta \left(t, \tau_i \mid \theta_i \right) = \omega \left( \tau_i \mid \theta \right)\) where \(\int_{0}^\infty \omega \left( \tau \mid \theta \right) d \tau = 1\) is the generation interval density. This is to say that all index cases share the same generation interval density, which is a function of the time since infection \(\tau_i\) parameterised by \(\theta\). The parameters \(\theta\) are specified, a priori, based on separate analyses of generation intervals for the epidemic in question. With these assumptions in place, \(p_{ji}\) can be computed for each index case as required. Letting \(R_j\) denote the expected number of secondary infections for the \(j^{th}\) index case, we have that
\[
R_j = \sum_{t_i > t_j} p_{ji},
\]
given that \(p_{ji} = 0\) if case \(i\) is known to have been imported. Finally, \(R_t\) is estimated by the arithmetic mean of \(R_j\) for all index cases in the cohort at time \(t\). The \wt{} method is implemented within the \texttt{EpiEstim} R package \citep{cori2020epiestim, RLanguage}.

\cite{bertozzi2020challenges} developed an approach to estimating \(R_t\) which extends \wt{}. Allowing for imported cases with intensity \(\mu\), they assume that \(\beta \left(t, \tau_i \mid \theta_i \right) = \mathcal R \left( t_i \right) \omega \left( \tau_i \mid \theta \right)\) such that \(\mathcal R \left( t \right) \equiv R_t\) and \(\omega \left( \tau_i \mid \theta \right)\) is the generation interval density adopted in \wt{}. Substituting these quantities into equations (\ref{eq:p_ji}) and (\ref{eq:p_ii}) provides the required relative likelihoods. \cite{bertozzi2020challenges} propose an iterative expectation-maximisation approach to maximum likelihood estimation for \(\mathcal R \left( t \right)\). To this end, they adopt a histogram estimator for \(\mathcal R \left( t \right)\) such that
\begin{equation}
    \mathcal R \left( t \right) = \sum_{k = 1}^{B} r_k \mathbbm 1 \left\{ t \in I_k \right\},
    \label{eq:bertozzi_Rt}
\end{equation}
where \(I_1, \dots I_B\) are a set of disjoint intervals that must be specified a priori and $\mathbbm 1 \left\{ t \in I_k \right\}$ is the usual indicator function taking the value $1$ if $t \in I_K$ and $0$ otherwise. Each interval defines a distinct cohort for which a constant value $r_k$ is estimated as 
\begin{equation}
    r_k = \frac{1}{N_k} \sum_{t_i > t_j} p_{ji} \mathbbm{1} \left \{ t_j \in I_k \right\},
    \label{eq:bertozzi_rk}
\end{equation}
where \(N_k\) is the total number of index cases on the interval \(I_k\). Thus, initialising \(\mathcal R \left( t \right)\) and iteratively updating equations (\ref{eq:p_ji}), (\ref{eq:bertozzi_Rt}), and (\ref{eq:bertozzi_rk}) allows \citeauthor{bertozzi2020challenges} to estimate the cohort reproduction number \(R_t\).

The methods presented above represent the current state-of-the-art approaches to estimating \(R_t\) within a branching process framework. Based on simple, parsimonious models which minimise the assumptions we must make, they offer informative results even with limited data.
However, the models proposed by \wt{} and \cite{bertozzi2020challenges} could be considered as restrictive in the following sense. Consider the number of secondary infections associated with an index case at time \(t\), which we denote by \(Z_t\). \wt{} implies that \(Z_t \sim \operatorname{Pois} \left( 1 \right)\) while \cite{bertozzi2020challenges} assume that \(Z_t \sim \operatorname{Pois} \left( R_t \right)\). 
This is a consequence of the definition of \(\beta \left(t, \tau_i \mid \theta_i \right)\) as the intensity function of a Poisson process. This observation reveals that neither model allows for reproduction numbers that vary from one individual to the next within each cohort. 
Adopting a hierarchical model for \(Z_t\) offers a framework for tackling this problem.
\cite{lloyd2005superspreading} provide a useful starting point, which assumes that
\begin{align*}
    Z_{t} \mid \nu_{t} &\sim \operatorname{Pois} \left( \nu_{t} \right), \\
    \nu_{t} &\sim \operatorname{Gamma} \left( \alpha, \beta \right), 
\end{align*}
where a Gamma distributed \textit{individual reproduction number} \(\nu_t\), parameterised by shape $\alpha$ and rate $\beta$, allows for over-dispersed secondary infections and heterogeneous disease reproduction. Defining $\alpha$ and $\beta$ in terms of the reproduction number $R$ and a dispersion parameter $k$, such that $\alpha = k$ and $\beta = k / R$, implies that $Z_{t} \sim \operatorname{Neg-Bin} \left( R, k \right)$, where \(\mathbb E \left[ Z_t \right] = R\) and \(\operatorname{Var} \left( Z_t \right) = R + R^2 / k\).

The continuous latent variable \(\nu_t\) models the complex combination of factors governing the infectiousness of each individual, assuming that infectiousness and susceptibility to infection are uncorrelated. Extending this framework to the time-varying reproduction number \(R_t\) offers an approach to modelling heterogeneity within branching process epidemic models.

A second issue is that empirical data on epidemics are not typically available in continuous time. Instead, the disease incidence is generally reported as a daily case count, i.e. at a series of discrete-time steps. The methods developed by \cite{wallinga2004different} or \cite{bertozzi2020challenges} are readily applied to such data; however, making the discrete-time formulation explicit allows for more efficient inference, given that computation of relative likelihoods (\ref{eq:p_ji}) and (\ref{eq:p_ii}) for all \(i, j\) scales with \(\mathcal O \left( N^2\right)\). When daily cases number in the tens to hundreds of thousands, this can present a significant computational burden, particularly if we are to extend these methods to heterogeneous disease reproduction.

Here, we extend these branching process epidemic models via a discrete-time model for epidemic curves where disease reproduction is heterogeneous within cohorts. This model retains the appealing parsimony of \cite{wallinga2004different} and \cite{bertozzi2020challenges} while allowing us to investigate the impact of heterogeneity. We develop a Bayesian approach to inference, applying state-of-the-art sampling techniques to provide a coherent approach to uncertainty quantification for $R_t$ and $k$ \citep{carpenter2017stan}.

\section{Methods}
\label{sec:methods}

\subsection{A generative model for epidemic curves}
\label{sec:generative}

Let \(Y_t\) denote the number of index cases recorded on day \(t = \dots, -1, 0, 1, 2, \dots\) such that the index cases up to and including day \(0\) seed the epidemic. 
We distinguish between imported and locally generated index cases, and so
\begin{equation*}
    Y_t = Y_t^{\operatorname{imp}} + Y_t^{\operatorname{loc}}.
    \label{eq:total_cases}
\end{equation*}
We let the number of cases imported on day \(t\) follow a Poisson distribution parameterised by rate \(\mu_t\), such that
\begin{equation*}
    Y_t^{\operatorname{imp}} \mid \mu_t \sim \operatorname{Pois} \left( \mu_t \right).
    \label{eq:imported_cases}
\end{equation*}
This implies that the imported case count is conditionally independent of those generated locally, given $\mu_t$.

In order to model the local incidence of disease \(Y_t^{\operatorname{loc}}\), that is the number of secondary infections recorded within the population on day \(t\), we let \(Z_{t, i}\) denote the number of secondary infections arising from the \(i^{th}\) index case on day \(t\). 
Adopting a hierarchical model for heterogeneous disease reproduction \citep{lloyd2005superspreading}, we have
\begin{align}
    Z_{t, i} \mid \nu_{t, i} &\sim \operatorname{Pois} \left( \nu_{t, i}\right),\label{eq:secondary_cases}\\
    \nu_{t, i} &\sim \operatorname{Gamma} \left( \alpha = k, \beta_t = \frac{k}{R_t} \right), \label{eq:reproductive_rate}
\end{align}
where the Gamma shape $\alpha$ and time-varying rate $\beta_t$ are defined in terms of the daily cohort reproduction number $R_t$ and dispersion parameter $k$.
The total reproduction number for all index cases on day \(t\) is then
\begin{equation}
\eta_{t} = \sum_{i = 1}^{Y_{t}}  \nu_{t, i} \sim \operatorname{Gamma} \left( Y_t k, \frac{k}{R_t} \right).
\label{eq:momentum}
\end{equation}
Adopting the nomenclature proposed by \cite{johnson2020disease}, we refer to the latent variable \(\eta_t\) as the disease momentum on day \(t\). The disease momentum describes the total infectiousness of all index cases recorded at time \(t\) when we have heterogeneous individual reproduction numbers, and defines the rate at which the epidemic spreads through the population.
That \(\eta_t\) is Gamma distributed follows from the fact that the sum of independent Gamma random variables with a common rate parameter is itself Gamma distributed.

\begin{figure}
    \centering
    \begin{tikzpicture}
    \tikzstyle{rv}=[circle, minimum size = 14mm, thick, draw = black!100, node distance = 21mm]
    \tikzstyle{tmp}=[node distance = 21mm]
    \tikzstyle{connect}=[-latex, thick]
    \tikzstyle{plate}=[rectangle, draw=black!100, thick, rounded corners=.5em]
    \node[rv, fill = black!20] (y) {\(Y_t\)};
    \node[tmp, above of = y] (tmp_above) {};
    \node[tmp, below of = y] (tmp_below) {};
    \node[tmp, below of = tmp_below] (tmp_bb) {};
    \node[rv, left of = tmp_above] (y_imp) {\(Y_t^{\operatorname{imp}}\)};
    \path (y_imp) edge [connect] (y);
    \node[rv, above of = y_imp] (mu) {\(\mu_t\)};
    \path (mu) edge [connect] (y_imp);
    \node[rv, left of = tmp_below] (y_loc) {\(Y_t^{\operatorname{loc}}\)};
    \path (y_loc) edge [connect] (y);
    \node[tmp, below of = y_loc] (tmp_local) {};
    \node[rv, left of = tmp_local] (eta_tms) {\(\eta_{t-s}\)};
    \path (eta_tms) edge [connect] (y_loc);
    \node[rv, left of = y_loc] (w_s) {\(\omega_{s}\)};
    \path (w_s) edge [connect] (y_loc);
    \node[tmp, left of = w_s] (tmp_w) {};
    \node[rv, above of = tmp_w, fill = black!20] (y_tms) {\(Y_{t-s}\)};
    \path (y_tms) edge [connect] (eta_tms);
    \node[rv, left of = eta_tms] (r_tms) {\(R_{t-s}\)};
    \path (r_tms) edge [connect] (eta_tms);
    \node[rv, right of = tmp_bb] (eta_t) {\(\eta_{t}\)};
    \path (y) edge [connect] (eta_t);
    \node[rv, left of = eta_t] (r_t) {\(R_{t}\)};
    \path (r_t) edge [connect] (eta_t);
    \node[rv, below of = tmp_local, node distance = 28mm] (k) {\(k\)};
    \path (k) edge [connect] (eta_t);
    \path (k) edge [connect] (eta_tms);
    \node[plate, fit= (y_tms) (eta_tms), label={[shift = {(0mm, -7.5mm)}]\(s = 1, 2, \dots\)}, inner sep = 5mm, yshift = 2.5mm] {};
    \node[plate, fit= (r_tms) (mu) (eta_t), label={[shift = {(0mm, -12.5mm)}]\(t = 1, 2, \dots\)}, inner sep = 10mm, yshift =  5mm] {};

    \end{tikzpicture}
    \caption{A plate diagram of the conditional dependence structure within the generative model for epidemic curves described in section \ref{sec:generative}, where only the epidemic curve \(\dots, Y_{-1}, Y_0, Y_1 \dots\) (shaded nodes) is observed. This figure highlights model parameters that are non-identifiable from the epidemic curve alone. Even if the disease momentum \(\eta_t\) were observed, joint inference for \(R_t\) and \(k\) depends on prior assumptions restricting the day-to-day variation in reproduction numbers. 
    }
    \label{fig:plates}
\end{figure}
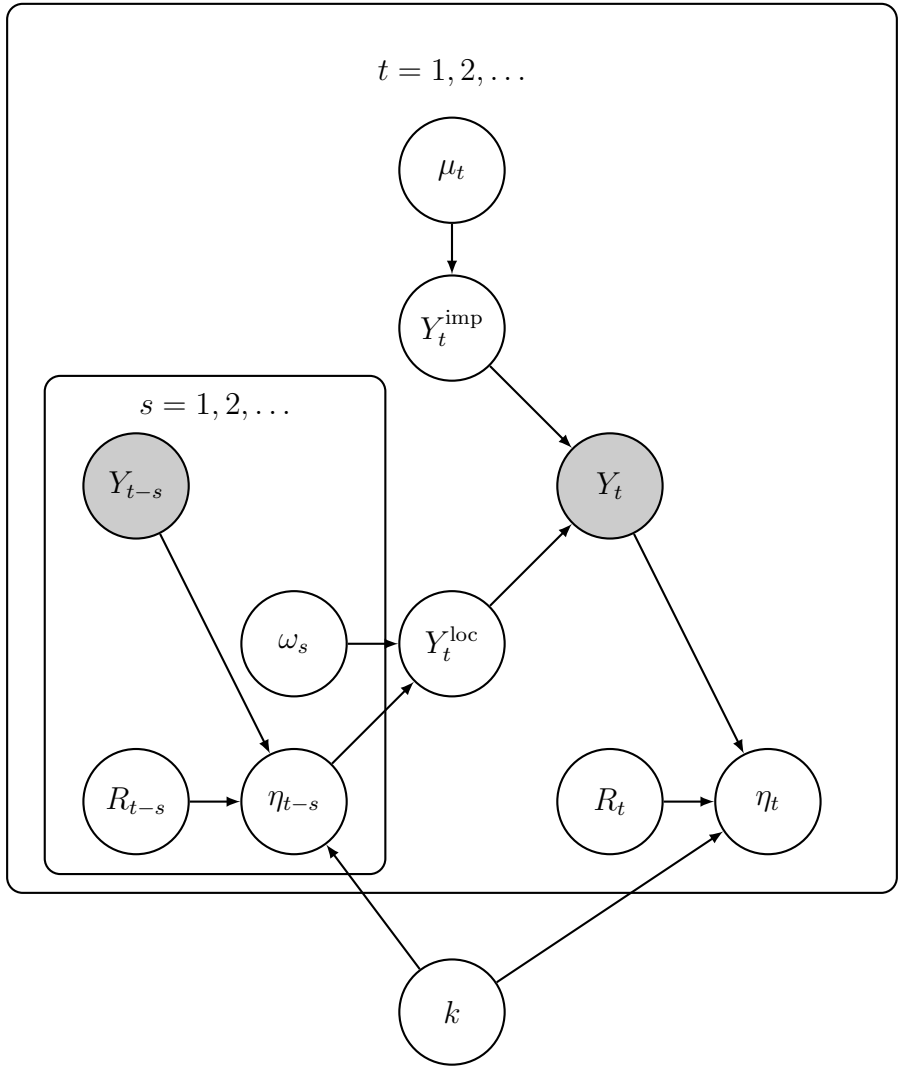

Given the generation interval probability mass function (pmf) \(\boldsymbol \omega = \left( \omega_1, \omega_2, \dots \right)^\top\), the distribution of secondary infections in time is modelled as a set of independent Poisson random variables. For the \(i^{th}\) index case on day \(t\), the number of secondary infections arising \(s\) days after infection can be expressed as
\[
Z_{t, i}^{t+s} \mid \nu_{t, i}, \boldsymbol \omega \sim \operatorname{Pois} \left( \omega_s \nu_{t, i} \right).
\]
This ensures that the model for secondary infections in equation (\ref{eq:secondary_cases}) holds.
Locally generated cases on day \(t\) are simply the sum over all secondary infections arising on day \(t\) from existing index cases, that is 
\[
Y_t^{\operatorname{loc}} = \sum_{s = 1}^{\infty} \sum_{i = 1}^{Y_{t-s}} Z_{t - s, i}^{t}.
\]
Thus, given \(\boldsymbol \omega\) and the disease momentum up to time \(t\), denoted \(\boldsymbol \eta_t = \left(\eta_{t-1}, \eta_{t-2}, \dots \right)^\top\), locally generated cases are Poisson distributed such that
\begin{equation}
    Y_t^{\operatorname{loc}} \mid \boldsymbol \omega, \boldsymbol \eta_{t} \sim \operatorname{Pois} \left(\sum_{s = 1}^{\infty} \omega_s \eta_{t-s}\right).
    \label{eq:local_cases}
\end{equation}
Adding imported cases to those generated locally, the likelihood for this generative model is
\begin{equation}
    Y_t  \mid \mu_t, \boldsymbol \omega, \boldsymbol \eta_{t} \sim \operatorname{Pois} \left(\mu_t + \sum_{s = 1}^{\infty} \omega_s \eta_{t-s}\right).
    \label{eq:full_likelihood}
\end{equation}
Figure \ref{fig:plates} presents a graphical representation of the dependence structure within this model for epidemic curves.

\subsubsection{Model identifiability}
\label{sec:identifiability}

Here we outline  issues relating to the identifiability of the model. In particular, suppose that we allow $\boldsymbol \omega$ to be a unknown parameter in the model.  

It can be shown that for any $\boldsymbol \omega' \neq \boldsymbol \omega$ there exists a unique $\boldsymbol \eta_t' \neq \boldsymbol \eta_t$ satisfying the linear equations defined by
$$
\sum_{s = 1}^{\infty} \omega_s' \eta_{t-s}' = \sum_{s = 1}^{\infty} \omega_s \eta_{t-s}.
$$
This implies, following equation (\ref{eq:full_likelihood}), that jointly inferring $\boldsymbol \omega$ and $\boldsymbol \eta_t$ leads to a non-identifiable model. For this reason we specify a fixed $\boldsymbol \omega$ for our model.
In Section~\ref{sec:data} we discuss our choice of $\boldsymbol \omega$ based on information drawn from the literature on generation intervals in the context of COVID-19.
Similarly, inferring both $\mu_t$ and $\boldsymbol \eta_t$ without observing $Y_t^{\operatorname{imp}}$ also leads to a non-identifiable model. 
To see this, suppose we have some constant $0 < c < \mu_t$, Then
$$
\mu_t' + \sum_{s = 1}^{\infty} \omega_s \eta_{t-s}' = \mu_t + \sum_{s = 1}^{\infty} \omega_s \eta_{t-s},
$$
is the case where $\mu_t' = \mu_t - c$ and $\eta_{t-s}' = \eta_{t - s} + c$. This shows that subtracting a constant from the import rate and adding it to each momentum variable leads to a non-identifiable likelihood in equation (\ref{eq:full_likelihood}).
For this reason we also specify a fixed $\mu_t$, as discussed in Section~\ref{sec:data}.

\subsubsection{Expected proportions of secondary infections} \label{sec:prop_sec}

The model for individual reproduction numbers in equations (\ref{eq:secondary_cases}) and (\ref{eq:reproductive_rate}) allows us to estimate the expected proportion of secondary infections associated with the most infectious index cases, as described by \cite{lloyd2005superspreading}. 

We consider index cases within each cohort separately, such that a single cohort reproduction number parameterises the distribution of individual reproduction numbers.
That is, we estimate the expected proportion of secondary infections associated with the most infectious individuals within the \(y_t\) index cases recorded on day \(t\), for whom the cohort reproduction number is \(R_t\). 
In this case, we model
\begin{equation}
    u_{t, i} = \frac{\nu_{t, i}}{R_t} \sim \operatorname{Gamma} \left( k, k \right),
\end{equation}
without any loss of generality. That is, within each cohort, the degree of heterogeneity in the distribution of individual reproduction numbers depends on \(k\) only.

Given \(p \left( x \mid \alpha, \beta \right)\) and \(F \left( x \mid \alpha, \beta \right)\), the probability density and cumulative distribution functions of a Gamma distributed random variable parameterised shape \(\alpha\) and rate \(\beta\), we define the cumulative distribution function for transmission of the disease within a single cohort as
\begin{equation}
\begin{aligned}
    F_{\operatorname{trans}} \left( x \mid k \right) &= \int _0^x u \, p \left( u \mid k, k \right) du, \\
    &= \frac{k^k}{\Gamma \left( k \right)} \int _0^x u^{k} \exp \left(- k u \right) du, \\
    &= \frac{k^k}{\Gamma \left( k \right)} \frac{\Gamma \left( k + 1 \right)}{k^{k+1}}\int _0^x p \left( u \mid k + 1, k \right) du, \\
    &= F \left( x \mid k + 1, k \right),
    \label{eq:transmission_percentile}
\end{aligned}
\end{equation}
where we have the identity \(\Gamma \left( k + 1 \right) = k \Gamma \left( k \right)\).
Here, \(F_{\operatorname{trans}} \left( x \mid k \right)\) is the expected proportion of secondary infections from index cases recorded day \(t\) attributable to individuals with \(u_{t, i} < x\).
The expected proportion of secondary infections due to individuals with \(u_{t, i} > x\) is therefore \(1 - F_{\operatorname{trans}} \left( x \mid k \right)\), while the proportion of individuals with \(u_{t, i} > x\) is \(1 - F \left( x \mid k, k \right)\) \citep{lloyd2005superspreading}.
Thus, if \(T_k  \left( q \right)\) denotes the expected proportion of secondary infections from the proportion \(q \in \left[ 0, 1 \right]\) of index cases within a single cohort that are most infectious, then \(T_k  \left( q \right) = 1 - F_{\operatorname{trans}} \left( x_q \mid k \right)\) where \(x_q\) satisfies \(1 - F \left( x_q \mid k, k \right) = q\). Written as a single expression, we have that
\begin{equation}
    T_k  \left( q \right) = 1 - F_{\operatorname{trans}} \left( F^{-1} \left( 1 - q \mid k, k \right) \mid k \right).
    \label{eqn:expected_transmission}
\end{equation}
As no closed form expression for \(F^{-1} \left( 1 - q \mid k, k \right)\) exists, we estimate \(T_k  \left( q \right)\) numerically for a given \(q\) and \(k\).

\subsection{Bayesian Analysis}

The generative model presented above offers a framework for learning about the spread of disease within a population given the observed epidemic curve \(\boldsymbol y = \left(y_{0}, y_1, \dots, y_N \right)^\top\) and a set of prior beliefs.
In the following, we present a Bayesian approach to inference, assuming that \(y_t^{\operatorname{imp}}\), the number of imported cases on day \(t = 1, \dots, N\), is unknown. Following the previous subsection, we assume that both the rate at which cases are imported \(\boldsymbol \mu = \left( \mu_0, \mu_1, \dots, \mu_N \right)^\top\) and the generation interval pmf $\boldsymbol \omega$ are fixed, a priori. In the following subsection, we outline how we specify a prior distribution for the daily cohort reproduction numbers $\boldsymbol R = \left(R_0, R_1, \dots, R_N\right)^\top$ and dispersion parameter $k$. For ease of exposition, we assume that the epidemic is seeded by \(y_0\) only, although in practice we include \(y_{-(N_0 - 1)}, \dots, y_{-1}\) to seed the epidemic with \(N_0\) days.

\subsubsection{Prior specification} \label{sec:priors}

We first address prior specification for the dispersion parameter $k$. 
Ideally, a prior for $k$ would rely on detailed contact tracing data which reconstructs the underlying transmission networks \citep{lloyd2005superspreading, arinaminpathy2020quantifying, sun2021transmission}; however, this information is often unavailable.
Our approach exploits the relationship between $k$ and the expected proportion of disease transmission attributable to the most infectious individuals, as set out in Section~\ref{sec:prop_sec}. Suppose we assume, for example, that the most infectious 20\% of index cases give rise to at least 30\% of expected secondary infections. By equation (\ref{eqn:expected_transmission}), this implies that $k < 10$. Similarly, if we assume that no more than 95\% of expected secondary infections arise from the most infectious 20\% of index cases, then equation (\ref{eqn:expected_transmission}) implies that $k > 0.1$. Thus, if our prior belief is that the most infectious 20\% of index cases give rise to 30-95\% of expected secondary infections, this leads to a prior distribution for $k$ with positive support in the interval $(0.1, 10)$. Note that this interval covers both high and low levels of heterogeneous disease transmission within the population. In addition, we note that for a simple SIR compartmental model, secondary infections follow a Geometric distribution, which corresponds to the case where $k = 1$ \citep{lloyd2005superspreading}. This suggests that a prior distribution with some central tendency towards $1$ is sensible and leads us to propose a log-Normal prior for $k$, such that
$$
\log k \sim \mathcal N \left( \mu_{\log k}, \sigma_{\log k}^2 \right),
$$
and set $\mu_{\log k} = 0$ and $\sigma_{\log k} = 1$. Under this prior, presented in terms of $T_k \left( q \right)$ in Figure \ref{fig:prior_transmission_quantile}, the median for $k$ is 1 and the same proportion of prior density is assigned to the interval $(0.1, 1)$ as is to $(1, 10)$. This prior distribution can be easily adapted to a more concentrated range of values for $k$ as required.

\textbf{\begin{figure}[t]
    \centering
    \includegraphics[width = 0.9\textwidth]{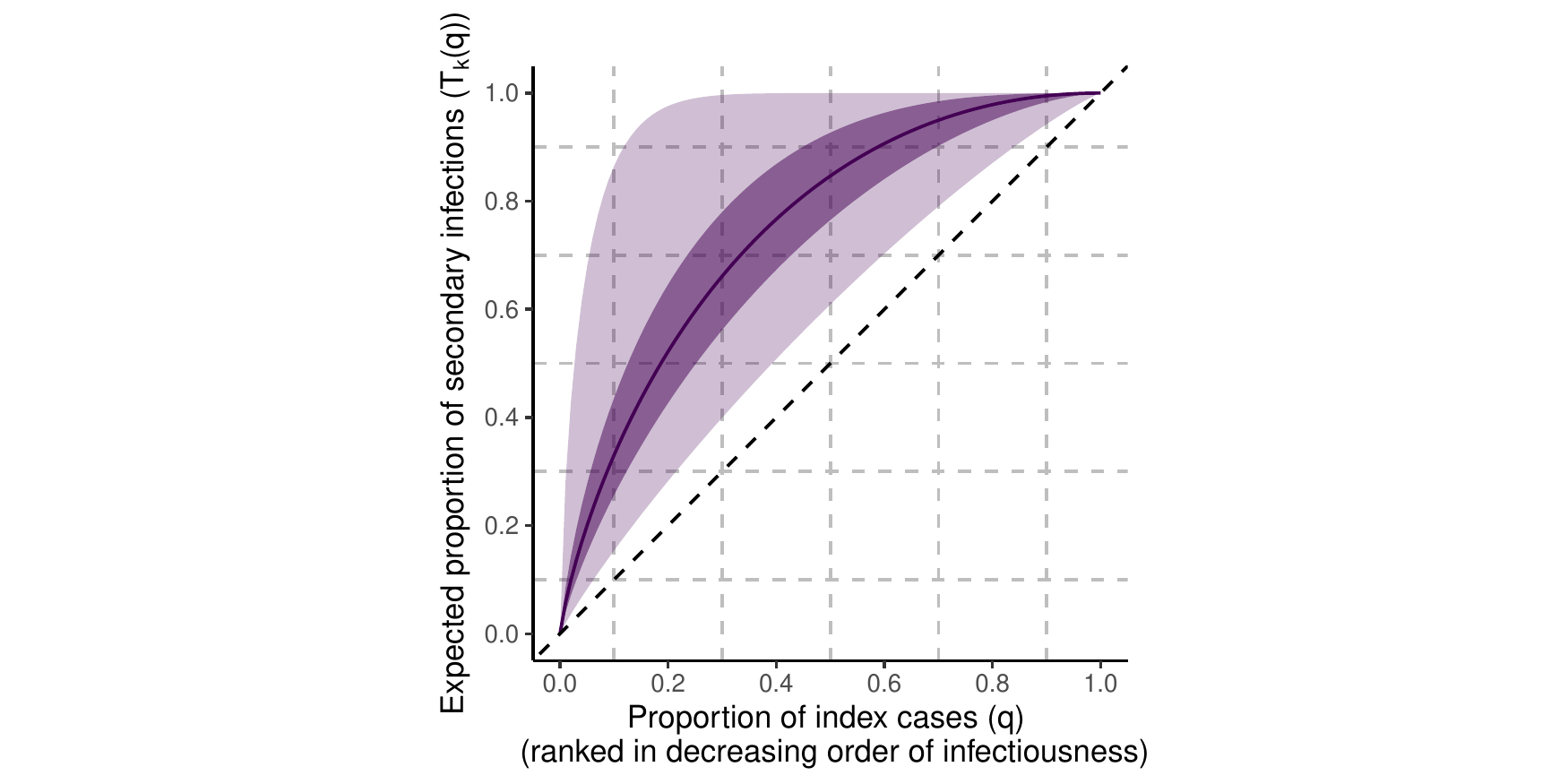}
    \caption{The distribution of the expected proportion of secondary infections arising from the most infectious proportion of individuals in the population under a $\log k \sim \mathcal N \left( 0, 1 \right)$ prior for $k$, summarised by the the prior median (solid line), inter-quartile range (dark shaded region), and 99\% equal-tailed interval (light shaded region). As an example, we see that this prior on $k$ provides support for 30-95\% of expected secondary infections arising from the most infectious 20\% of index cases.}
    \label{fig:prior_transmission_quantile}
\end{figure}}

The cohort reproduction number \(R_t\) is defined in equations (\ref{eq:secondary_cases}) and (\ref{eq:reproductive_rate}) as the expected number of secondary infections arising from each index case in the cohort recorded on day \(t\).
As described by \cite{fraser2007estimating} and \cite{gostic2020practical}, this expectation should vary smoothly in time. For example, if control measures introduced at \(t_c\) restrict the transmission of disease, some cohorts infected at time \(t < t_c\) will spend part of their infectious period both before and after the introduction of control measures. Thus, \(R_t\) will transition from high to low values smoothly at a rate which depends on \(\boldsymbol \omega\).

We model \(R_t\) as a smooth function of time via a Gaussian process (GP) prior \citep{williams2006gaussian}. Given that \(R_t > 0\), let
\begin{equation}
    R_t = \exp \left( f \left( t \right) \right),
    \label{eq:link_f_Rt}
\end{equation}
where the latent function \(f \left( t \right) \in \mathbb R\) is a GP such that
\begin{equation}
    \begin{aligned}
    f \left( t \right) &\sim \mathcal{GP} \left( 0, k \left( t, t' \right) \right), \\
    k \left( t, t' \right) & = \sigma_f^2 \exp \left( - \frac{\left( t - t' \right)^2}{2 \ell^2} \right),
    \end{aligned}
    \label{eq:gp_prior}
\end{equation}
with amplitude \(\sigma_f > 0\) and length-scale \(\ell > 0\). 
The exponentiated quadratic covariance covariance function \(k \left( \cdot, \cdot \right)\) implies that \(f \left( t \right)\) is infinitely differentiable, where \(\sigma_f\) defines the expected range of \(f \left( \cdot \right)\) and \(\ell\) governs the functions rate of change, which we loosely describe as its ``wiggliness''.

Our approach to specifying a prior distribution for $\boldsymbol R$ 
involves relating the expected number of zero-upcrossings of $f \left( \cdot \right)$ 
to the expected number of surges in case numbers per year, as we now outline. 
If we assume that \(R_t\) is a slowly-varying function such that sustained increases in daily case numbers occur when \(R_t > 1\), sustained decreases imply that \(R_t < 1\), and short term variation is driven by heterogeneous disease reproduction, then distinct surges in case numbers are associated with \(R_t\) crossing \(1\) from below. 
Thus, given that \(f \left( t \right) = 0\) when \(R_t = 1\), each surge is associated with a zero-upcrossing of \(f \left( \cdot \right)\). 
If we let \(\mathbb E \left[ n_0 \right]\) denote the expected number of zero-upcrossings of \(f \left( \cdot \right)\) 
per year, then it can be shown for (\ref{eq:gp_prior}) that \(\mathbb E \left[ n_0 \right] = 365*\left( 2 \pi \ell \right)^{-1}\) \citep{williams2006gaussian}, where the length-scale is in units of days. 
Thus, we can relate the number of surges in case numbers we expect over a given period to \(\ell\).
For example, if we expect 3--4 surges in case numbers per year, then $\ell$ should lie on the interval \(\left(15, 20\right)\). This insight allows us to specify a prior distribution for $\ell$, which we outline in Section~\ref{sec:data}. We refer the reader to Appendix \ref{app:prior} for further details on prior elicitation for the GP.
 
When inferring cohort reproduction numbers for an ongoing epidemic, a full Bayesian treatment of \(\ell\) might not be practical, due to the computational cost incurred (evaluation of the Gaussian density scales with \(\mathcal O \left( N^3 \right)\)). In this case, $\ell$ may be fixed, a priori. Inference for $\boldsymbol R$ is not particularly sensitive to the value chosen for $\ell$, although smaller values tend to inflate estimates for $k$. 
The approach which we take subsequently is to place a zero-truncated Gaussian hyper-prior on $\ell$ such that $\ell \sim \mathcal N \left( \mu_\ell, \sigma_\ell^2 \right)$. 
Note that we are typically only interested values for $\ell$ that are far from 0, and so the effect of truncation on this hyper-prior can be safely ignored.

\subsubsection{Posterior Inference}

Our primary objective is to learn about the joint posterior over $\boldsymbol R$ and $k$ given the epidemic curve $\boldsymbol y$. To do this, we integrate over the unknown momentum variables $\boldsymbol \eta = \left( \eta_0, \eta_1, \dots, \eta_N \right)^\top$, latent GP $\boldsymbol f = \left(f_0, f_1, \dots, f_N \right)^\top$, and the GP length-scale $\ell$. As detailed in Section~\ref{sec:identifiability},  we treat $\boldsymbol \mu$ and $\boldsymbol \omega$ as fixed parameters, and specify the hyper-parameters $\sigma_f, \mu_{\log k}, \sigma_{\log k}, \mu_{\ell}, \sigma_{\ell}$. Thus, we wish to infer the marginal posterior distribution
$$
p \left(\boldsymbol R, k \mid \boldsymbol y \right) = \int p \left(\boldsymbol R, k, \boldsymbol \eta, \boldsymbol f, \ell \mid \boldsymbol y \right) d \boldsymbol \eta \, d\boldsymbol f \, d\ell.
$$
The joint posterior distribution of all unknown quantities can be written as
\begin{equation}
p \left(\boldsymbol R, k, \boldsymbol \eta, \boldsymbol f, \ell \mid \boldsymbol y \right) \propto 
p \left( \boldsymbol y, \boldsymbol \eta \mid \boldsymbol R, k \right) 
p \left(\boldsymbol R, \boldsymbol f \mid \ell \right) 
p \left( \ell \right) p \left( k \right),
\label{eq:joint_conditional}
\end{equation}
where the complete-data likelihood is expressed as
\begin{equation}
    p \left( \boldsymbol y, \boldsymbol \eta \mid \boldsymbol R, k \right) = \prod_{t = 1}^{N} p \left( \eta_{t-1} \mid y_{t-1}, R_{t-1}, k \right) p \left(y_t \mid \boldsymbol \eta_t \right). 
    \label{eq:complete_data}
\end{equation}
For ease of expression we have suppressed notation conditioning on the fixed parameters and hyper-parameters. Note that this epidemic model is seeded by \(y_0\) and so these cases are omitted from the likelihood in equation (\ref{eq:complete_data}).

This hierarchical model for \(\boldsymbol y\) is defined by
\begin{equation}
\begin{aligned}
y_t \mid \boldsymbol \eta_t, \mu_t, 
\boldsymbol \omega  &\sim \operatorname{Pois} \left(\mu_t + \sum_{s = 1}^t \omega_s \eta_{t - s} \right), \\
\eta_{t} \mid y_{t}, R_{t}, k  &\sim \operatorname{Gamma} \left( y_t \, k, \frac{k}{R_t}\right), \\
\log k &\sim \mathcal N \left( \mu_{\log k}, \sigma_{\log k}^2 \right), \\
\boldsymbol R &= \exp \left( \boldsymbol f \right),  \\
\boldsymbol f \mid \sigma_f, \ell  &\sim \mathcal N \left( \boldsymbol 0, \boldsymbol K \right), \\
\ell &\sim \mathcal N \left( \mu_\ell, \sigma_\ell^2 \right).
\end{aligned}
\label{eq:bayesian_model}
\end{equation}
where \(\boldsymbol K\) is the Gram matrix of the covariance function in (\ref{eq:gp_prior}). The probabilistic model described above is coded in Stan \citep{carpenter2017stan, rstan, RLanguage} and can be implemented using the 
R package {\texttt{assessEpidemicCurves}} which can found at \url{https://github.com/jpmeagher/assessEpidemicCurves}.

\section{Data Description and Model Specification}
\label{sec:data}

We consider the COVID-19 epidemic in Ireland as a case study. Note that an alternative analysis of COVID-19 in Ireland, developed by the Irish Epidemiological Modelling Advisory Group (IEMAG) to inform the Irish Government's response to the epidemic, is presented by \cite{gleeson2022calibrating}. 
We assess the 7-day moving average of confirmed cases, ordered by epidemiological date. 
The epidemiological date is the earliest recorded date associated with a confirmed case of COVID-19.
This is either the date of onset of symptoms, date of diagnosis, the laboratory specimen collection date, the laboratory received date, the laboratory reported date or the notification date.
Sorting cases by their epidemiological date strips out some random effects on the epidemic curve introduced by reporting delays, while taking the 7-day moving average of case counts smooths over other day-of-the-week effects. 
This data was extracted from the Computerised Infectious Disease Reporting database hosted by the Health Protection Surveillance Centre (HPSC) and provided to the authors by the Central Statistics Office of Ireland.

\begin{figure}[t]
    \centering
    \includegraphics[width = 0.9 \textwidth]{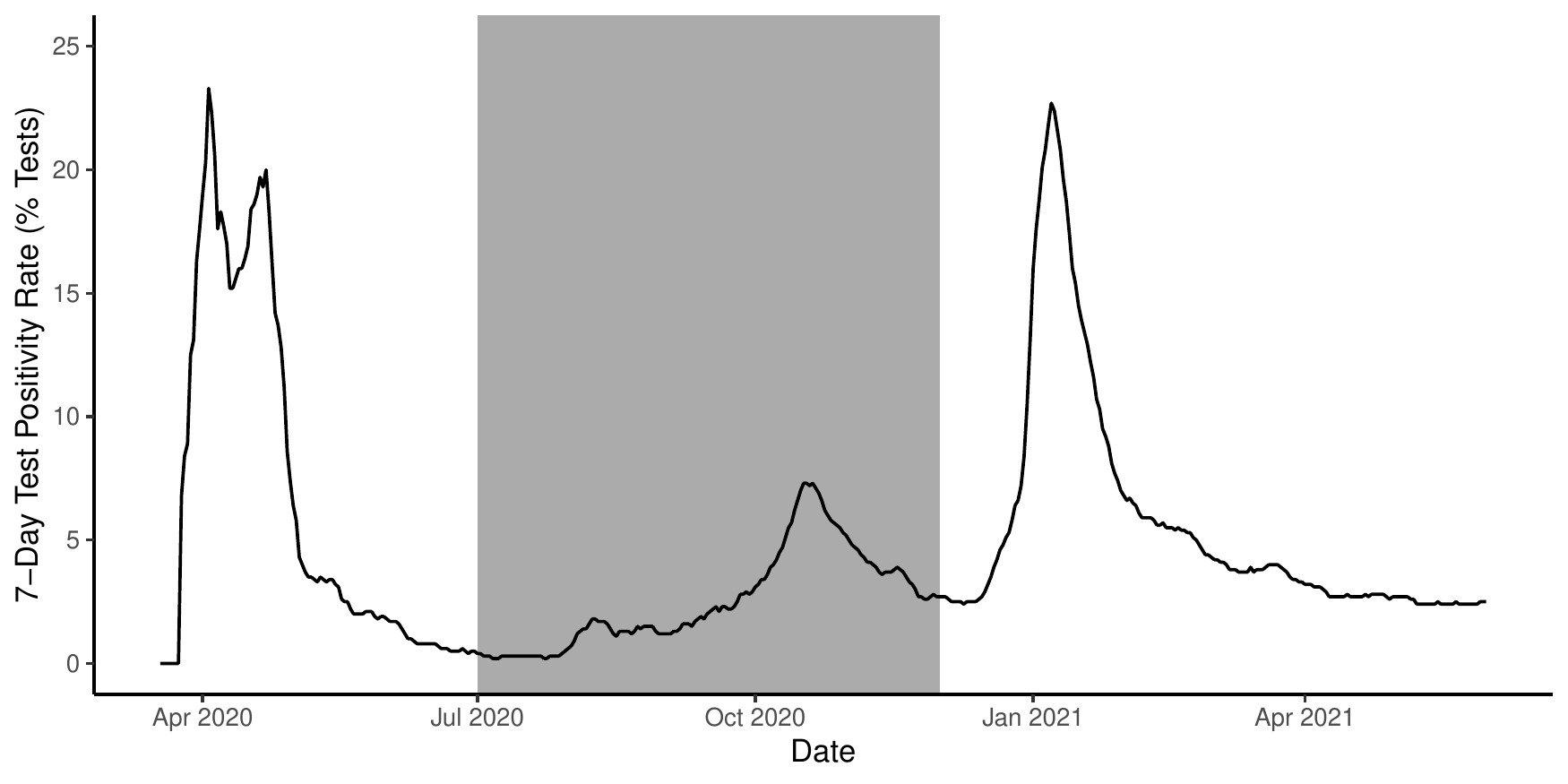}
    \caption{The 7-day test positivity rate as a percentage of all tests reported by the HPSC up to June 2021. The shaded region covers the time period included in our analysis. Note that the test positivity never exceeds 10\% in this period while it exceeds 20\% at in April 2020 and January 2021. Our assumption is that a high positivity rate is indicative of a testing system that has been overwhelmed by cases, resulting in less reliable daily case counts.}
    \label{fig:positivity}
\end{figure}

Our analysis focuses on the period from July to November 2020, covering Ireland's second surge in coronavirus infections up to the easing of restrictions in December. We have restricted our analysis to this period as we believe that the Irish testing and contact tracing system functioned in a relatively consistent manner throughout. Figure \ref{fig:positivity} presents the 7-day average test positivity rate reported by the HPSC up to June 2021. We see that the positivity rate did not exceed 10\% in our analysis period (the shaded region in the figure), unlike during the first surge in March/April 2020 and the third surge in December 2020/January 2021. This suggests that testing capacity was better able to cope with demand in this period, which we expect to result in more consistent testing and tracing procedures. Thus, we expect recorded data from this period to reflect the ongoing epidemic more accurately than at other points in time, while still recording a growing epidemic. Our assumption is that consistent testing and tracing procedures will allow us to more accurately assess the epidemic curve for evidence of superspreading.

Estimates for the daily count of imported cases \(\boldsymbol y^{\operatorname{imp}}\) are not available for this dataset. As such, we simply assume that the rate at which cases are imported is 1 for all \(t\) such that \(\boldsymbol \mu\) is a vector of ones. This assumption may be unrealistic. We might expect that $\mu_t$ depends on the incidence of COVID-19 across the United Kingdom and European Union, for example, and the rate at which individuals travel between Ireland and these jurisdictions. However, given the absence of relevant data, our assumption seems reasonable. A small value for $\mu_t$ implies that the vast majority of cases are due to local transmission and provides a similar approach to that taken by the IEMAG, where imported cases were omitted from the model entirely \citep{gleeson2022calibrating}. An additional, practical consideration is that non-zero values for $\mu_t$ avoid numerical issues within our inference scheme, although seeding the epidemic with $\mu_0 \approx y_0$ is generally sufficient to avoid any difficulties.

Letting $\gamma_\tau$ and $\sigma_\tau$ denote a mean and standard deviation for the distribution of generation intervals,
we assume that \(\boldsymbol \omega\) follows a discretised Gamma distribution with mean \(\gamma_\tau = 5\) and standard deviation \(\sigma_\tau = 2.5\), truncated at $S = 21$ days. This is achieved by setting
\begin{equation}
    \begin{aligned}
    \omega_1 &\propto \int_{0}^{1.5} \operatorname{Gamma} \left( x \mid \alpha = \left( \frac{\gamma_\tau}{\sigma_\tau}\right)^2, \beta = \frac{\gamma_\tau}{\sigma_\tau^2} \right) dx, &\\
    \omega_s &\propto \int_{s-0.5}^{s + 0.5} \operatorname{Gamma} \left( x \mid \alpha = \left( \frac{\gamma_\tau}{\sigma_\tau}\right)^2, \beta = \frac{\gamma_\tau}{\sigma_\tau^2} \right) dx, &\text{for } s = 2, \dots, S,
    \label{eq:generation_interval_pmf}
    \end{aligned}
\end{equation}
such that \(\sum_{s = 1}^S \omega_s = 1\), where we have matched $\gamma_\tau$ and $\sigma_\tau$ to the shape \(\alpha\) and rate \(\beta\) of the Gamma distribution.
In the absence of detailed contact tracing information for infector-infectee pairs in Ireland, these values for $\gamma_\tau$ and $\sigma_\tau$ have been chosen from the middle of the range for COVID-19 reported in the literature \citep{du2020serial, ganyani2020estimating, griffin2020rapid, rai2020estimates}. 
In addition, truncating $\boldsymbol \omega$ at $21$ days implies that the maximum generation interval is three weeks, an assumption is reasonably consistent with empirical data \citep{du2020serial}.
Although our analysis considers a single, fixed parameterisation for $\boldsymbol \omega$, the posterior distribution is quite robust to changes in $\gamma_\tau$ and $\sigma_\tau$. In Appendix \ref{app:gen_int}, we consider models parameterised by $(\gamma_\tau, \sigma_\tau) \in \left\{ (4, 2), (6, 3)\right\}$, which represent lower and upper estimates of $\gamma_\tau$ respectively, as reported by \cite{rai2020estimates}. This analysis shows that each model provides broadly similar inference for $\boldsymbol R$ and $k$. 

We specify the GP prior for \(\boldsymbol R\) with hyper-parameters \(\sigma_f = 1\), \(\mu_\ell = 17.5\) and $\sigma_\ell = 2.5$. Under this prior, the marginal prior distribution for \(\log R_t\) is normally distributed with mean \(0\) and standard deviation 1. Setting $\sigma_f = 1$ implies that the 95\% credible interval for \(R_t\) under this marginal prior distribution is \(\left( 0.14, 7.10 \right)\) with mean \(1.65\) and a median at \(1\). This marginal distribution covers the range of values we expect $R_t$ to take, a priori. Assuming that \(\ell \sim \mathcal N \left( 17.5, 2.5^2 \right)\), left truncated at zero, implies that we expect 2--5 surges in case numbers per year. Given that Ireland experienced 3 surges over the first year of its coronavirus epidemic, this hyper-prior is not overly restrictive and covers a reasonable range of possible values for $\ell$. See Appendix \ref{app:prior} for further analysis supporting this hyper-prior distribution.
Finally, we remind the reader that we have proposed a weakly informative prior for $k$ in Section \ref{sec:priors}, specified as $\log k \sim N(\mu_{\log k},\sigma_{\log k}^2)$ where $\mu_{\log k} = 0$ and $\sigma_{\log k} = 1$.

\section{Analysis and Results}
\label{sec:results}

We present our analysis of the Irish COVID-19 epidemic curve as follows. We first fit the model defined by (\ref{eq:bayesian_model}) and specified in Section \ref{sec:data}, evaluate the fit of our model via the posterior predictive distribution, and assess the epidemic curve for evidence of heterogeneous disease transmission. Following this, we compare the posterior inference for $\boldsymbol R$ to those obtained under the assumption of homogeneous disease transmission.

\subsection{Heterogeneous disease transmission} \label{sec:hetero}

We fit our model to the Irish epidemic curve for the $N = 150$ days from 4 July to 30 November 2020, inclusive, allowing the $N_0 = 5$ days up to 4 July to seed the epidemic. We draw 4 chains of 5000 samples from (\ref{eq:joint_conditional}) after a warm-up of 2000 samples and thin our chains by retaining every fifth sample. Posterior samples satisfy standard diagnostic checks and tests for convergence \citep{vehtari2021rank}.

\begin{figure}[t]
    \centering
    \includegraphics[width = 0.9\textwidth]{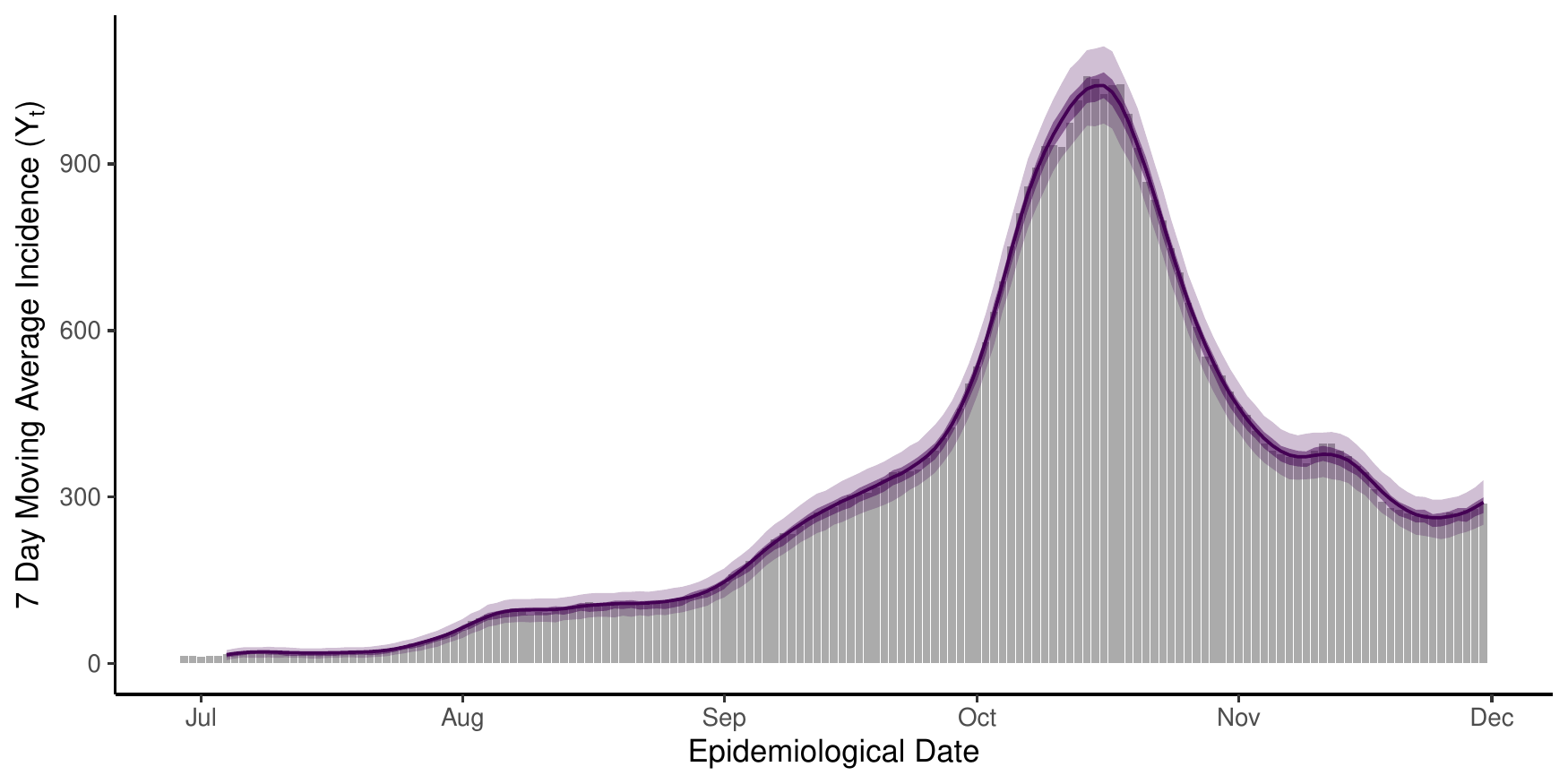}
    \caption{The mean (solid line), 50\% (dark shaded region), and 95\% credible interval (light shaded region) for the posterior predictive distribution of the COVID-19 epidemic curve in Ireland. Note that the predictive model is seeded by \(N_0 = 5\) days. All observed daily case counts fall within the 95\% credible interval of the posterior predictive distribution with 86\% of counts within the 50\% credible interval.}
    \label{fig:pp_figure}
\end{figure}

The posterior predictive distribution for our model is
$$
p \left( \tilde{\boldsymbol{y}} \mid \boldsymbol y \right) = \int p \left( \tilde{\boldsymbol{y}} \mid \boldsymbol \eta, \boldsymbol y \right) p \left( \boldsymbol \eta \mid \boldsymbol R, k, \boldsymbol y \right) p \left( \boldsymbol R, k \mid y \right) d \boldsymbol \eta \, d \boldsymbol R \, dk,
$$ 
which we obtain by integrating over the sampled posterior distribution of the momentum variables. In Figure \ref{fig:pp_figure} we see that the posterior predictive distribution tracks the empirical epidemic curve closely and provides good coverage of daily cases counts. All observed daily counts fall within the 95\% credible interval while 86\% fall within the 50\% credible interval. This reasonably well balanced posterior predictive distribution suggests that our model offers a good fit to the data.

\begin{figure}[t]
\centering
\begin{subfigure}{0.475\textwidth}
    \includegraphics[width=\textwidth]{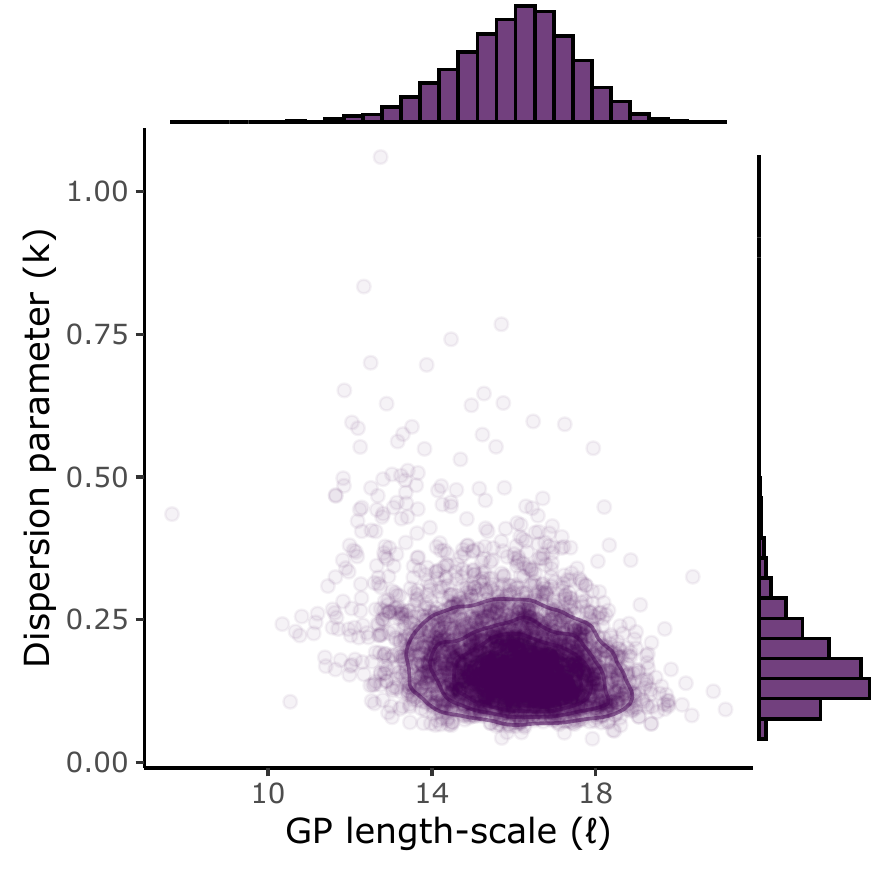}
    \caption{}
    \label{fig:lk_figure}
\end{subfigure}
\begin{subfigure}{0.475\textwidth}
    \includegraphics[width=\textwidth]{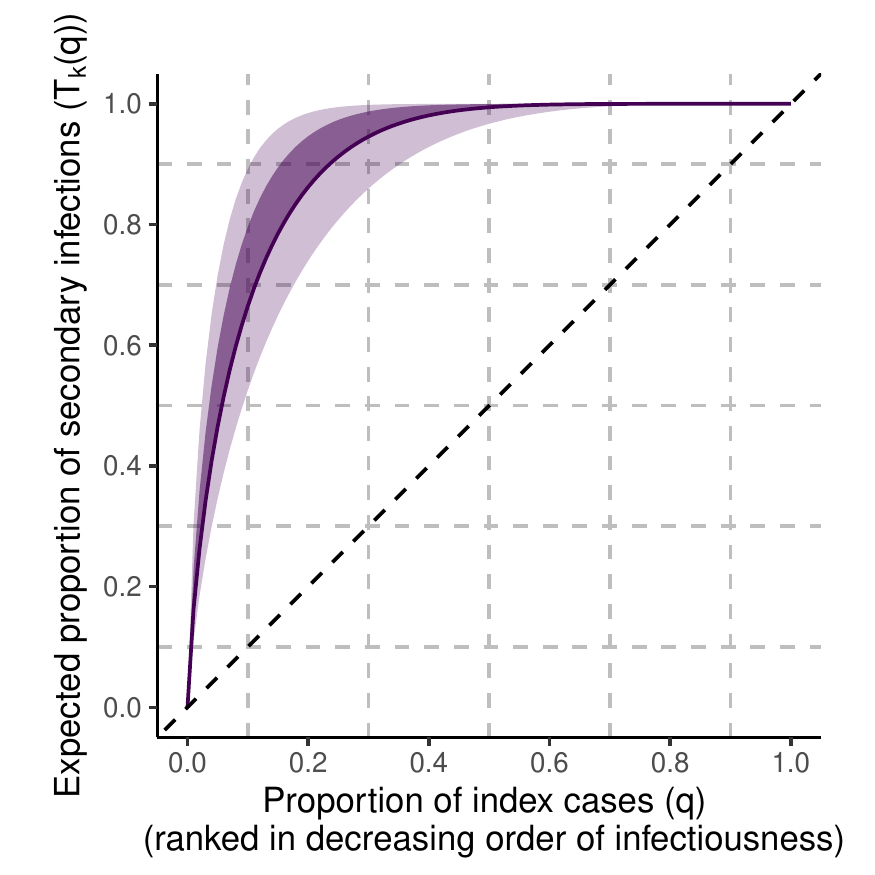}
    \caption{}
    \label{fig:Tq_figure}
\end{subfigure}
        
\caption{(a) The sampled joint and marginal posterior distributions over $k$ and $\ell$. Each opaque point represents a sample, while contours are set to have a width of 0.3. We report 95\% credible intervals for $k$ of $\left( 0.07, 0.33 \right)$ and for $\ell$ of $\left( 13, 18 \right)$. 
(b) The proportion of expected secondary infections attributable to the proportion $q$ of most infectious individuals. The mean (solid line), 50\% (dark shaded region), and 95\% credible interval (light shaded region) for this proportion over the interval $q \in \left[ 0, 1\right]$ is presented. $T_k \left( q \right)$ is estimated numerically by equation (\ref{eqn:expected_transmission}) given the posterior distribution for $k$. Based on this analysis we estimate, for example, that the 20\% most infectious individuals give rise to 75--98\% of expected secondary infections with $95\%$ posterior probabililty, while 62--82\% of individuals are not expected to pass on the infection, again with $95\%$ posterior probability.}
\label{fig:hetero_figure}
\end{figure}

Figure \ref{fig:lk_figure} presents the sampled marginal and joint posterior distributions over the dispersion parameter $k$ and GP length-scale hyper-parameter $\ell$. This figure illustrates that there is a weak inverse relationship between $\ell$ and $k$, with large values for $\ell$ (i.e. very slowly varying cohort reproduction numbers) tending to be associated with smaller values for $k$ (i.e. more heterogeneous disease transmission). Based on this analysis, we infer 95\% credible intervals of approximately $\left( 0.07, 0.33 \right)$ for $k$ and $\left( 13, 18 \right)$ for $\ell$.

The uncertainty in $k$, a posteriori, allows us assess the posterior uncertainty in the estimated proportion of expected secondary infections associated with the proportion $q$ of most infectious individuals, using the approach developed in Section~\ref{sec:prop_sec}. To do this, we sample from the posterior distribution of $k$ and solve (\ref{eqn:expected_transmission}) numerically for each sampled value. This gives rise to a distribution for the expected proportion of secondary infections for a given proportion $q$. We present this for $q \in \left[0, 1 \right]$ in Figure~\ref{fig:Tq_figure}. If we consider the 20\% most infectious individuals as an example, this analysis suggests that 75--98\% of expected secondary infections can be attributed to these individuals with 95\% posterior probability, while 62--82\% of infected individuals are not expected to pass on the infection, again with 95\% posterior probability. This represents a high degree of heterogeneity in the spread of COVID-19.

\subsection{Time-varying reproduction numbers}

Given that our model provides evidence for heterogeneous disease transmission, our next objective is to compare inference for \(R_t\) under this model with those assuming homogeneous disease reproduction. We adapt the model defined in (\ref{eq:bayesian_model}) to homogeneous disease reproduction by fixing $\eta_t = R_t y_t$ for all $t$. In effect, this assumes that $k \to \infty$. As above, we draw 4 chains of 5000 samples after a warm-up of 2000 samples and retain every fifth sample to thin our chains. Once again, posterior samples satisfy our diagnostic checks and tests for convergence.
In addition, we fit \citeauthor{wallinga2004different}' (\citeyear{wallinga2004different}) model (\wt{}) to our epidemic curve using the \texttt{EpiEstim} package \citep{cori2020epiestim}. When estimating and quantifying uncertainty on \(R_t\) by \wt{}, the cohort at time \(t\) is defined by the 3-day window such that the estimate for \(R_t\) is the arithmetic mean number of secondary infections arising from index cases on days \(t-1, t,\) and \(t+1\). 

This analysis demonstrates that heterogeneous disease reproduction has important consequences when estimating \(R_t\). While estimates of the posterior mean show a general agreement across all three approaches, credible intervals around these estimates behave very differently when we assume heterogeneous disease reproduction within cohorts. Empirically, we observe that credible intervals under homogeneous disease transmission and \wt{} tend to be only 60\% as wide as those under heterogeneous disease transmission at time \(t\). If our objective was to establish whether or not \(R_t \neq 1\), then allowing for heterogeneous disease transmission is a crucial consideration.

It is also worth noting that, under homogeneous disease transmission, the GP length-scale hyper-parameter $\ell$ has a 95\% credible interval of $\left(7, 13 \right)$, despite the $\ell \sim \mathcal N \left( 17.5, 2.5 \right)$ hyper-prior distribution. This inference reflects the fact that homogeneous disease transmission requires a flexible model for $\boldsymbol R$ in order to fit empirical data well. This model for $\boldsymbol R$ implies that we should expect 5-8 surges in case numbers per year, far more than we observe empirically.

\begin{figure}[t]
    \centering
    \includegraphics[width = 0.9\textwidth]{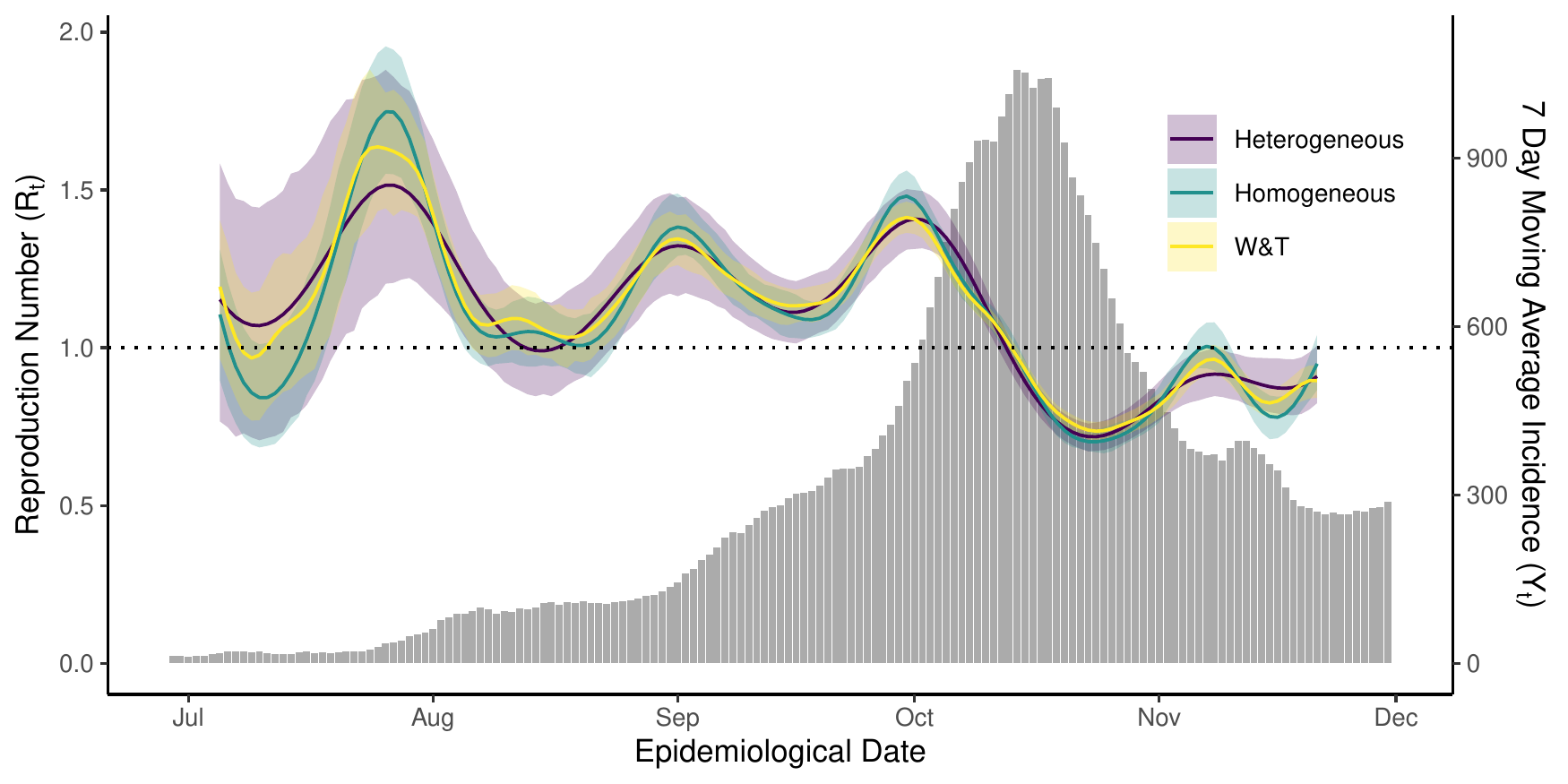}
    \caption{The posterior mean (solid line) and 95\% credible interval (shaded region) for \(\boldsymbol R\) under the heterogeneous, homogeneous, and \wt{} model for disease transmission. At time points well supported by data, estimates for \(R_t\) provided by each of the three methods show a general agreement. However, credible intervals under \wt{} or the assumption of homogeneous disease reproduction within each cohort are approximately 60\% as wide as those in the heterogeneous case at each point in time. In addition, note that our estimate for $R_t$ is less ``wiggly'' under heterogeneous disease transmission than in the homogeneous case. This behaviour illustrates that a more flexible model for $\boldsymbol R$ is required to fit data when we assume homogeneous disease transmission.}
    \label{fig:Rt_figure}
\end{figure}

\section{Discussion}
\label{sec:discussion}

In this report, we have developed a parsimonious model for epidemic curves in the presence of heterogeneous disease reproduction, offering a Bayesian extension to the work of \cite{wallinga2004different} and \cite{bertozzi2020challenges}. This model treats superspreading as a feature of the epidemic rather than a phenomenon that occurs rarely. We develop a Bayesian inference scheme based on a GP prior for the cohort reproduction number \(R_t\) and a log-Normal prior on the dispersion parameter $k$. This hierarchical model allows us to assess the degree of heterogeneity in individual reproduction numbers supported by any given epidemic curve, providing insight into the distribution of secondary infections within the population.
An R package implementing these methods is available at \href{https://github.com/jpmeagher/assessEpidemicCurves}{\texttt{github.com/jpmeagher/assessEpidemicCurves}}.

Our analysis of the COVID-19 epidemic in Ireland provides support for heterogeneous disease reproduction.
This result, alongside mounting evidence from other jurisdictions (see, e.g. \cite{endo2020estimating, arinaminpathy2020quantifying, sun2021transmission}), leads us to conclude that superspreading is a salient feature of this epidemic. 
A useful output of our analysis is that it allows us to estimate the expected proportion of secondary infections attributable to the most infectious proportion of the population. For example, we estimate that the 20\% most infectious individuals give rise to 75--98\% of the expected secondary infections with 95\% posterior probability, while 62--82\% of individuals did not pass on the infection, also with 95\% posterior probability.
This finding has important implications for public health policy. 
In particular, our analysis suggests that heterogeneity should be accounted for when quantifying uncertainty on \(R_t\).
We observed that credible intervals for \(R_t\) are typically much wider when \(k\) is small than is the case for homogeneous disease reproduction. This uncertainty is an important consideration when deciding whether or not to implement public health interventions based on \(R_t\). Secondly, when disease reproduction is heterogeneous, control measures targeting the most infectious individuals will have a disproportionate impact on the overall disease momentum \citep{woolhouse1997heterogeneities, lloyd2005superspreading, wallinga2010optimizing}. In this instance, backward tracing, which looks to identify the source of each infection, could play a crucial role in bringing the epidemic under control while minimising the broader societal and economic impact.
Finally, the uncertainty introduced by heterogeneous disease transmission should be accounted for when forecasting the trajectory of an epidemic.

The proposed framework comes with important caveats. The parsimonious model ignores several features of empirical epidemic curves. In reality, the epidemic curve is never observed under ideal conditions, the rate at which cases are imported is unknown, and the distribution of generation intervals is likely to change in response to both control measures and the growth rate of the epidemic \citep{ali2020serial, park2021forward}. Misspecification of these parameters will effect the joint distribution over \(R_t\) and \(k\) \citep{wallinga2007generation, gostic2020practical, knight2020estimating, donnat2020modeling}. 
That said, in our analysis of the Irish COVID-19 epidemic, we have sought to mitigate the worst of these effects by modelling an epidemic curve where cases have been ordered by epidemiological date rather than considering the daily count of confirmed cases. We have also restricted our analysis to a period in which we have reason to believe that testing and tracing systems could cope with demand and procedures were executed in a reasonably consistent manner.

This paper should provide a starting point for research in several directions. For example, if transmission networks are available, then it would be possible to incorporate this type of branching process data into the framework developed here to provide enhanced inference on the heterogeneity of disease spread. 
Additionally, extending our model to allow joint inference over $\mu_t$, $R_t$, $k$ and $\boldsymbol \omega$ is also possible, if richer epidemic data are available.

\section*{Acknowledgements} 
The Insight Centre for Data Analytics is supported by Science Foundation Ireland under Grant Number 12/RC/2289$\_$P2.

\bibliographystyle{agsm.bst}
\bibliography{bibliography}

@article{ali2020serial,
  title={Serial interval of {SARS-CoV-2} was shortened over time by non-pharmaceutical interventions},
  author={Ali, Sheikh Taslim and Wang, Lin and Lau, Eric HY and Xu, Xiao-Ke and Du, Zhanwei and Wu, Ye and Leung, Gabriel M and Cowling, Benjamin J},
  journal={Science},
  volume={369},
  number={6507},
  pages={1106--1109},
  year={2020},
  publisher={American Association for the Advancement of Science}
}

@article{arinaminpathy2020quantifying,
  title={Quantifying heterogeneity in {SARS-CoV-2} transmission during the lockdown in {India}},
  author={Arinaminpathy, Nimalan and Das, Jishnu and McCormick, Tyler H and Mukhopadhyay, Partha and Sircar, Neelanjan},
  journal={Epidemics},
  volume={36},
  pages={100477},
  year={2021},
  publisher={Elsevier}
}

@book{anderson1992infectious,
  title={Infectious diseases of humans: dynamics and control},
  author={Anderson, Roy M and May, Robert M},
  year={1992},
  publisher={Oxford university press}
}

@article{becker1999statistical,
  title={Statistical studies of infectious disease incidence},
  author={Becker, Niels G and Britton, Tom},
  journal={Journal of the Royal Statistical Society: Series B (Statistical Methodology)},
  volume={61},
  number={2},
  pages={287--307},
  year={1999},
  publisher={Wiley Online Library}
}

@article{bertozzi2020challenges,
  title={The challenges of modeling and forecasting the spread of {COVID-19}},
  author={Bertozzi, Andrea L and Franco, Elisa and Mohler, George and Short, Martin B and Sledge, Daniel},
  journal={Proceedings of the National Academy of Sciences},
  volume={117},
  number={29},
  pages={16732--16738},
  year={2020},
  publisher={National Acad Sciences}
}

@article{bjornstad2020modeling,
  title={Modeling infectious epidemics},
  author={Bj{\o}rnstad, Ottar N and Shea, Katriona and Krzywinski, Martin and Altman, Naomi},
  journal={Nat. Methods},
  volume={17},
  number={5},
  pages={455--456},
  year={2020}
}

@article{britton2010stochastic,
  title={Stochastic epidemic models: a survey},
  author={Britton, Tom},
  journal={Mathematical biosciences},
  volume={225},
  number={1},
  pages={24--35},
  year={2010},
  publisher={Elsevier}
}

@article{britton2020mathematical,
  title={A mathematical model reveals the influence of population heterogeneity on herd immunity to {SARS-CoV-2}},
  author={Britton, Tom and Ball, Frank and Trapman, Pieter},
  journal={Science},
  volume={369},
  number={6505},
  pages={846--849},
  year={2020},
  publisher={American Association for the Advancement of Science}
}

@article{carpenter2017stan,
  title={Stan: A probabilistic programming language},
  author={Carpenter, Bob and Gelman, Andrew and Hoffman, Matthew D and Lee, Daniel and Goodrich, Ben and Betancourt, Michael and Brubaker, Marcus and Guo, Jiqiang and Li, Peter and Riddell, Allen},
  journal={Journal of statistical software},
  volume={76},
  number={1},
  year={2017},
  publisher={Columbia Univ., New York, NY (United States); Harvard Univ., Cambridge, MA~…}
}

@article{cori2013new,
  title={A new framework and software to estimate time-varying reproduction numbers during epidemics},
  author={Cori, Anne and Ferguson, Neil M and Fraser, Christophe and Cauchemez, Simon},
  journal={American journal of epidemiology},
  volume={178},
  number={9},
  pages={1505--1512},
  year={2013},
  publisher={Oxford University Press}
}

@Manual{cori2020epiestim,
    title = {EpiEstim: Estimate time-varying reproduction numbers from epidemic curves},
    author = {Anne Cori},
    year = {2020},
    note = {R package version 2.2-3},
    url = {https://CRAN.R-project.org/package=EpiEstim},
}

@article{dempster1977maximum,
  title={Maximum likelihood from incomplete data via the {EM} algorithm},
  author={Dempster, Arthur P and Laird, Nan M and Rubin, Donald B},
  journal={Journal of the Royal Statistical Society: Series B (Methodological)},
  volume={39},
  number={1},
  pages={1--22},
  year={1977},
  publisher={Wiley Online Library}
}

@article{donnat2020modeling,
  title={Modeling the heterogeneity in {COVID-19's} reproductive number and its impact on predictive scenarios},
  author={Donnat, Claire and Holmes, Susan},
  journal={Journal of Applied Statistics},
  pages={1--29},
  year={2021},
  publisher={Taylor \& Francis}
}

@article{du2020serial,
  title={Serial interval of {COVID-19} among publicly reported confirmed cases},
  author={Du, Zhanwei and Xu, Xiaoke and Wu, Ye and Wang, Lin and Cowling, Benjamin J and Meyers, Lauren Ancel},
  journal={Emerging infectious diseases},
  volume={26},
  number={6},
  pages={1341},
  year={2020},
  publisher={Centers for Disease Control and Prevention}
}

@article{endo2020estimating,
  title={Estimating the overdispersion in {COVID-19} transmission using outbreak sizes outside {China}},
  author={Endo, Akira and Abbott, Sam and Kucharski, Adam J and Funk, Sebastian and others},
  journal={Wellcome Open Research},
  volume={5},
  number={67},
  pages={67},
  year={2020},
  publisher={F1000 Research Limited}
}

@article{flaxman2020estimating,
  title={Estimating the effects of non-pharmaceutical interventions on {COVID-19} in {Europe}},
  author={Flaxman, Seth and Mishra, Swapnil and Gandy, Axel and Unwin, H Juliette T and Mellan, Thomas A and Coupland, Helen and Whittaker, Charles and Zhu, Harrison and Berah, Tresnia and Eaton, Jeffrey W and others},
  journal={Nature},
  volume={584},
  number={7820},
  pages={257--261},
  year={2020},
  publisher={Nature Publishing Group}
}

@article{fraser2007estimating,
  title={Estimating individual and household reproduction numbers in an emerging epidemic},
  author={Fraser, Christophe},
  journal={PloS one},
  volume={2},
  number={8},
  pages={e758},
  year={2007},
  publisher={Public Library of Science}
}

@article{forsberg2008likelihood,
  title={A likelihood-based method for real-time estimation of the serial interval and reproductive number of an epidemic},
  author={Forsberg White, Laura and Pagano, Marcello},
  journal={Statistics in medicine},
  volume={27},
  number={16},
  pages={2999--3016},
  year={2008},
  publisher={Wiley Online Library}
}

@article{ganyani2020estimating,
  title={Estimating the generation interval for coronavirus disease ({COVID-19}) based on symptom onset data, {March} 2020},
  author={Ganyani, Tapiwa and Kremer, C{\'e}cile and Chen, Dongxuan and Torneri, Andrea and Faes, Christel and Wallinga, Jacco and Hens, Niel},
  journal={Eurosurveillance},
  volume={25},
  number={17},
  pages={2000257},
  year={2020},
  publisher={European Centre for Disease Prevention and Control}
}

@article{gleeson2022calibrating,
  title={Calibrating COVID-19 susceptible-exposed-infected-removed models with time-varying effective contact rates},
  author={Gleeson, James P and Brendan Murphy, Thomas and O’Brien, Joseph D and Friel, Nial and Bargary, Norma and O'Sullivan, David JP},
  journal={Philosophical Transactions of the Royal Society A},
  volume={380},
  number={2214},
  pages={20210120},
  year={2022},
  publisher={The Royal Society}
}

@article{gostic2020practical,
  title={Practical considerations for measuring the effective reproductive number, ${R_t}$},
  author={Gostic, Katelyn M and McGough, Lauren and Baskerville, Edward B and Abbott, Sam and Joshi, Keya and Tedijanto, Christine and Kahn, Rebecca and Niehus, Rene and Hay, James A and De Salazar, Pablo M and others},
  journal={PLoS computational biology},
  volume={16},
  number={12},
  pages={e1008409},
  year={2020},
  publisher={Public Library of Science San Francisco, CA USA}
}

@article{grassly2008mathematical,
  title={Mathematical models of infectious disease transmission},
  author={Grassly, Nicholas C and Fraser, Christophe},
  journal={Nature Reviews Microbiology},
  volume={6},
  number={6},
  pages={477--487},
  year={2008},
  publisher={Nature Publishing Group}
}

@article{griffin2020rapid,
  title={Rapid review of available evidence on the serial interval and generation time of {COVID-19}},
  author={Griffin, John and Casey, Miriam and Collins, {\'A}ine and Hunt, Kevin and McEvoy, David and Byrne, Andrew and McAloon, Conor and Barber, Ann and Lane, Elizabeth Ann and More, Simon},
  journal={BMJ open},
  volume={10},
  number={11},
  pages={e040263},
  year={2020},
  publisher={British Medical Journal Publishing Group}
}

@article{hadeler1995core,
  title={A core group model for disease transmission},
  author={Hadeler, Karl P and Castillo-Ch{\'a}vez, Carlos},
  journal={Mathematical biosciences},
  volume={128},
  number={1-2},
  pages={41--55},
  year={1995},
  publisher={Elsevier}
}

@article{hoffman2014no,
  title={The {No-U-Turn} sampler: adaptively setting path lengths in {Hamiltonian Monte Carlo.}},
  author={Hoffman, Matthew D and Gelman, Andrew},
  journal={J. Mach. Learn. Res.},
  volume={15},
  number={1},
  pages={1593--1623},
  year={2014}
}

@article{johnson2020disease,
  title={Disease momentum: estimating the reproduction number in the presence of superspreading},
  author={Johnson, Kory D and Beiglb{\"o}ck, Mathias and Eder, Manuel and Grass, Annemarie and Hermisson, Joachim and Pammer, Gudmund and Polechov{\'a}, Jitka and Toneian, Daniel and W{\"o}lfl, Benjamin},
  journal={Infectious Disease Modelling},
  volume={6},
  pages={706--728},
  year={2021},
  publisher={Elsevier}
}

@book{keeling2011modeling,
  title={Modeling infectious diseases in humans and animals},
  author={Keeling, Matt J and Rohani, Pejman},
  year={2011},
  publisher={Princeton University Press}
}

@article{keeling2005networks,
  title={Networks and epidemic models},
  author={Keeling, Matt J and Eames, Ken TD},
  journal={Journal of the Royal Society Interface},
  volume={2},
  number={4},
  pages={295--307},
  year={2005},
  publisher={The Royal Society London}
}

@article{kermack1927contribution,
  title={A contribution to the mathematical theory of epidemics},
  author={Kermack, William Ogilvy and McKendrick, Anderson G},
  journal={Proceedings of the royal society of london. Series A, Containing papers of a mathematical and physical character},
  volume={115},
  number={772},
  pages={700--721},
  year={1927},
  publisher={The Royal Society London}
}

@article{knight2020estimating,
  title={Estimating effective reproduction number using generation time versus serial interval, with application to {COVID-19} in the {Greater Toronto Area, Canada}},
  author={Knight, Jesse and Mishra, Sharmistha},
  journal={Infectious Disease Modelling},
  volume={5},
  pages={889--896},
  year={2020},
  publisher={Elsevier}
}

@article{koopman2004modeling,
  title={Modeling infection transmission},
  author={Koopman, Jim},
  journal={Annu. Rev. Public Health},
  volume={25},
  pages={303--326},
  year={2004},
  publisher={Annual Reviews}
}

@article{lloyd2005superspreading,
  title={Superspreading and the effect of individual variation on disease emergence},
  author={Lloyd-Smith, James O and Schreiber, Sebastian J and Kopp, P Ekkehard and Getz, Wayne M},
  journal={Nature},
  volume={438},
  number={7066},
  pages={355--359},
  year={2005},
  publisher={Nature Publishing Group}
}

@article{may1987transmission,
  title={Transmission dynamics of {HIV} infection},
  author={May, Robert M and Anderson, Roy M},
  journal={Nature},
  volume={326},
  number={6109},
  pages={137--142},
  year={1987},
  publisher={Nature Publishing Group}
}

@article{park2021forward,
  title={Forward-looking serial intervals correctly link epidemic growth to reproduction numbers},
  author={Park, Sang Woo and Sun, Kaiyuan and Champredon, David and Li, Michael and Bolker, Benjamin M and Earn, David JD and Weitz, Joshua S and Grenfell, Bryan T and Dushoff, Jonathan},
  journal={Proceedings of the National Academy of Sciences},
  volume={118},
  number={2},
  year={2021},
  publisher={National Acad Sciences}
}

@article{rai2020estimates,
  title={Estimates of serial interval for {COVID-19}: A systematic review and meta-analysis},
  author={Rai, Balram and Shukla, Anandi and Dwivedi, Laxmi Kant},
  journal={Clinical epidemiology and global health},
  volume={9},
  pages={157--161},
  year={2021},
  publisher={Elsevier}
}

@Manual{RLanguage,
title = {R: A language and environment for statistical computing},
author = {{R Core Team}},
organization = {R Foundation for Statistical Computing},
address = {Vienna, Austria},
year = {2020},
url = {https://www.R-project.org/},
}

@Misc{rstan,
title = {{RStan}: the {R} interface to {Stan}},
author = {{Stan Development Team}},
note = {R package version 2.21.3},
year = {2020},
url = {http://mc-stan.org/},
}

@article{schmidt2020inference,
  title={Inference under superspreading: Determinants of {SARS-CoV-2} transmission in {Germany}},
  author={Schmidt, Patrick W},
  journal={arXiv preprint arXiv:2011.04002},
  year={2020}
}

@article{shen2004superspreading,
  title={Superspreading {SARS} events, {Beijing}, 2003},
  author={Shen, Zhuang and Ning, Fang and Zhou, Weigong and He, Xiong and Lin, Changying and Chin, Daniel P and Zhu, Zonghan and Schuchat, Anne},
  journal={Emerging infectious diseases},
  volume={10},
  number={2},
  pages={256},
  year={2004},
  publisher={Centers for Disease Control and Prevention}
}

@article{sun2021transmission,
  title={Transmission heterogeneities, kinetics, and controllability of {SARS-CoV-2}},
  author={Sun, Kaiyuan and Wang, Wei and Gao, Lidong and Wang, Yan and Luo, Kaiwei and Ren, Lingshuang and Zhan, Zhifei and Chen, Xinghui and Zhao, Shanlu and Huang, Yiwei and others},
  journal={Science},
  volume={371},
  number={6526},
  year={2021},
  publisher={American Association for the Advancement of Science}
}

@article{thompson2019improved,
  title={Improved inference of time-varying reproduction numbers during infectious disease outbreaks},
  author={Thompson, RN and Stockwin, JE and van Gaalen, Rolina D and Polonsky, JA and Kamvar, ZN and Demarsh, PA and Dahlqwist, Elisabeth and Li, Siyang and Miguel, Eve and Jombart, Thibaut and others},
  journal={Epidemics},
  volume={29},
  pages={100356},
  year={2019},
  publisher={Elsevier}
}

@article{veen2008estimation,
  title={Estimation of space-time branching process models in seismology using an {EM-type} algorithm},
  author={Veen, Alejandro and Schoenberg, Frederic P},
  journal={Journal of the American Statistical Association},
  volume={103},
  number={482},
  pages={614--624},
  year={2008},
  publisher={Taylor \& Francis}
}

@article{vehtari2021rank,
  title={Rank-normalization, folding, and localization: An improved {R} for assessing convergence of {MCMC}},
  author={Vehtari, Aki and Gelman, Andrew and Simpson, Daniel and Carpenter, Bob and B{\"u}rkner, Paul-Christian},
  journal={Bayesian analysis},
  volume={1},
  number={1},
  pages={1--28},
  year={2021},
  publisher={International Society for Bayesian Analysis}
}

@article{wallinga2004different,
  title={Different epidemic curves for severe acute respiratory syndrome reveal similar impacts of control measures},
  author={Wallinga, Jacco and Teunis, Peter},
  journal={American Journal of epidemiology},
  volume={160},
  number={6},
  pages={509--516},
  year={2004},
  publisher={Oxford University Press}
}

@article{wallinga2007generation,
  title={How generation intervals shape the relationship between growth rates and reproductive numbers},
  author={Wallinga, Jacco and Lipsitch, Marc},
  journal={Proceedings of the Royal Society B: Biological Sciences},
  volume={274},
  number={1609},
  pages={599--604},
  year={2007},
  publisher={The Royal Society London}
}

@article{wallinga2010optimizing,
  title={Optimizing infectious disease interventions during an emerging epidemic},
  author={Wallinga, Jacco and van Boven, Michiel and Lipsitch, Marc},
  journal={Proceedings of the National Academy of Sciences},
  volume={107},
  number={2},
  pages={923--928},
  year={2010},
  publisher={National Acad Sciences}
}

@book{williams2006gaussian,
  title={Gaussian processes for machine learning},
  author={Williams, Christopher KI and Rasmussen, Carl Edward},
  year={2006},
  publisher={MIT press Cambridge, MA}
}

@article{woolhouse1997heterogeneities,
  title={Heterogeneities in the transmission of infectious agents: implications for the design of control programs},
  author={Woolhouse, Mark EJ and Dye, C and Etard, J-F and Smith, T and Charlwood, JD and Garnett, GP and Hagan, P and Hii, JLK and Ndhlovu, PD and Quinnell, RJ and others},
  journal={Proceedings of the National Academy of Sciences},
  volume={94},
  number={1},
  pages={338--342},
  year={1997},
  publisher={National Acad Sciences}
}

@article{wong2020evidence,
  title={Evidence that coronavirus superspreading is fat-tailed},
  author={Wong, Felix and Collins, James J},
  journal={Proceedings of the National Academy of Sciences},
  volume={117},
  number={47},
  pages={29416--29418},
  year={2020},
  publisher={National Acad Sciences}
}

\newpage

\appendix

\section{Branching process transmission networks}
\label{app:branching}

The branching process model defined by equation (\ref{eq:conditional_intensity}) has the likelihood
\begin{equation}
p \left( \boldsymbol t \mid \boldsymbol \theta \right) = \prod_{i = 1}^N \lambda^* \left( t_i \right)  \exp \left( - \Lambda \left( T \right)\right), 
    \label{eq:liklihood}
\end{equation}
where \(\boldsymbol \theta = \left( \theta_1, \dots, \theta_N \right)\) and
\[
\Lambda \left( t \right) = \int_{0}^t \lambda^* \left( s \right) ds.
\]
This branching process model implies that there exists an underlying transmission network linking infector-infectee pairs where nodes represent individual cases and directed edges the chain of transmission.
Given that each secondary infection is associated with one index case only, each node in this transmission network has a single incoming edge. As such, the network can be uniquely characterised by the vector \(\boldsymbol v = \left( v_0, v_1, \dots v_N \right)^\top \in V\) where
\[
v_i = 
\begin{cases}
j, & \text{if case } i \text{ is a secondary infection of index case } j, \\
i, & \text{if case } i \text{ has been imported from outside the population,}
\end{cases}
\]
and \(V\) is the set of all possible transmission networks. The adjacency matrix \(\boldsymbol A\) corresponding to the transmission network \(\boldsymbol v\) is defined by \(\boldsymbol A_{ji} = \mathbbm 1 \left\{ v_i = j \right\}\) for \(j, i = 0, 1, \dots, N\), where \(\mathbbm 1 \left\{ \cdot \right\}\) is the indicator function. 
Thus, the complete-data likelihood for the branching process model can be expressed as
\begin{equation}
    p \left( \boldsymbol t, \boldsymbol A \mid \boldsymbol \theta \right) =  \left( \prod_{i = 1}^N \mu^{\boldsymbol A_{ii}} \prod_{j = 0}^{i-1} \beta \left(t_i, t_i - t_j \mid \theta_j \right)^{\boldsymbol A_{ji}} \right) \exp \left( - \Lambda \left( T \right)\right).
    \label{eq:complete_data_liklihood}
\end{equation}

The transmission network of an epidemic is generally unknown; however, by the chain rule of probability, we have that the ratio of (\ref{eq:complete_data_liklihood}) to (\ref{eq:liklihood}) yields a conditional distribution for \(\boldsymbol A\) such that
\[
p \left( \boldsymbol A \mid \boldsymbol t, \boldsymbol \theta \right) = \prod_{i = 1}^N  \frac{\mu^{\boldsymbol A_{ii}} \prod_{j = 0}^{i-1} \beta \left(t_i, t_i - t_j \mid \theta_j \right)^{\boldsymbol A_{ji}}}{\lambda^* \left( t_i \right)}.
\]
This probability mass function (pmf) is equivalent to the product of independent multinomial pmfs 
\[
\boldsymbol a_{i} \mid \boldsymbol t, \boldsymbol \theta \sim \operatorname{Multinomial} \left( 1, \boldsymbol p_i \right)
\]
where, for \(i = 1, \dots, N\), \(\boldsymbol a_{i} = \left( \boldsymbol A_{0i}, \boldsymbol A_{1i}, \dots, \boldsymbol A_{ii} \right)^\top\) is the outcome vector identifying the source of index case \(i\) and \(\boldsymbol p_i = \left( p_{0i}, p_{1i}, \dots, p_{ii} \right)^\top \) is a vector of probabilities such that
\begin{equation*}
    p_{ji} = \frac{\beta \left(t_i, t_i - t_j \mid \theta_j \right)}{\lambda^* \left( t_i \right)},
    \label{eq:app_p_ji}
\end{equation*}
is the relative likelihood that \(i\) is a secondary infection of \(j\) for all \(j = 0, \dots, i-1\) and
\begin{equation*}
    p_{ii} = \frac{\mu}{\lambda^* \left( t_i \right)},
    \label{eq:app_p_ii}
\end{equation*}
is the relative likelihood that \(i\) was imported.
Thus, the distribution of edges in the transmission network follows a series of independent trials with multinomial outcomes. 

\section{GP prior elicitation}
\label{app:prior}

The choice of hyper-parameters in the GP prior for the time-varying reproduction number is a crucial aspect of the analysis presented here. The prior for \(\boldsymbol R = \left( R_0, R_1, \dots, R_N \right)^\top\), defined by equations (\ref{eq:link_f_Rt}) and (\ref{eq:gp_prior}), is parameterised by amplitude \(\sigma_f > 0\) and length-scale \(\ell > 0\). As discussed in the main text, specification of \(\ell\) is an important consideration.

\begin{figure}
    \centering
    \includegraphics[width = 0.9\textwidth]{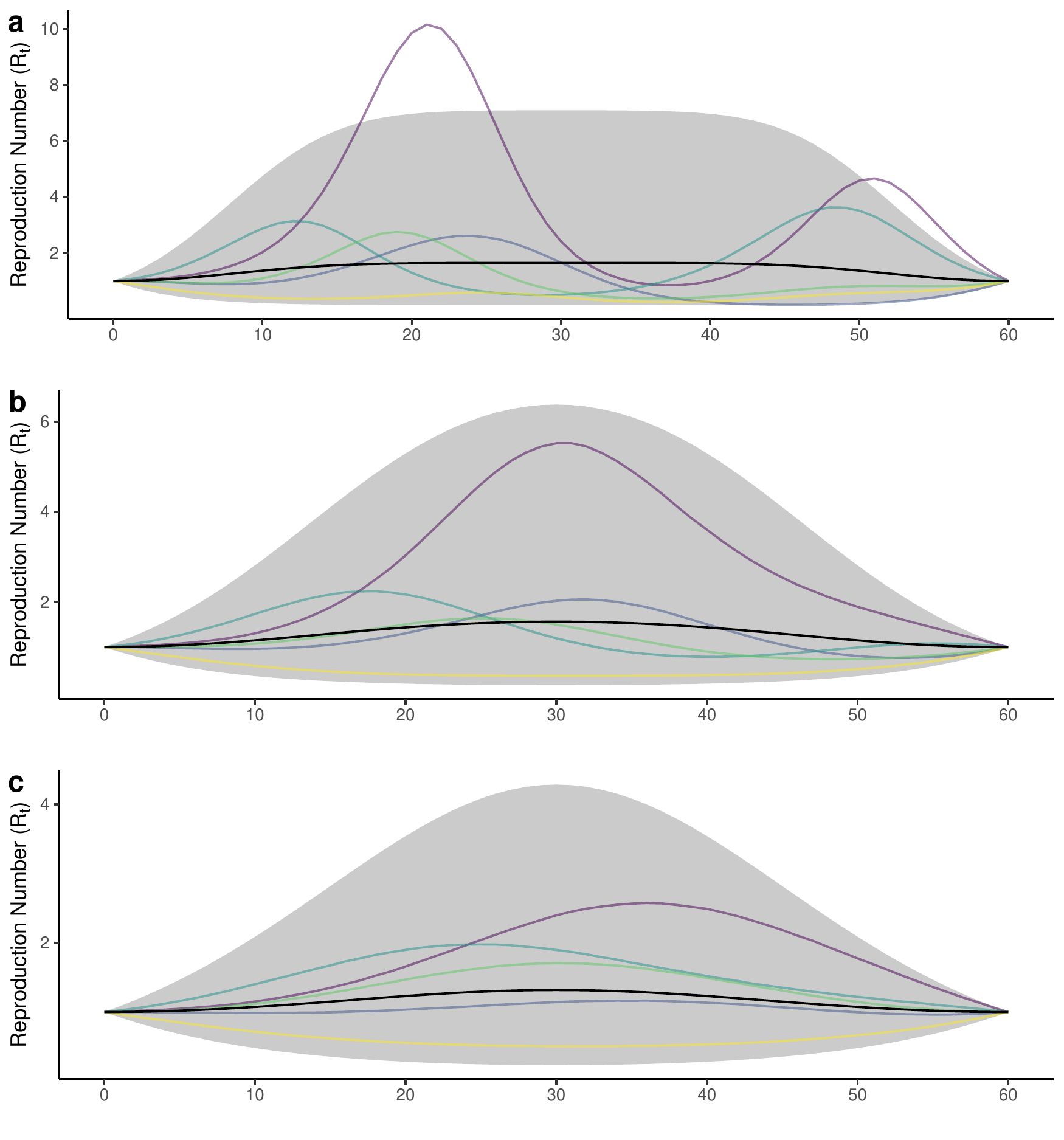}
    \caption{Summaries of the distribution for \(\boldsymbol R\) under a GP prior conditional on \(R_0 = R_N = 1\) and \(\sigma_f = 1\) where \(T = 60\). Three prior specifications are considered: \(\ell = 10\) (\textbf{a}); \(\ell = 17.5\) (\textbf{b}); and \(\ell = 25\) (\textbf{c}). In each case the solid black line represents the conditional mean alongside the shaded 95\% credible interval. Coloured lines represent samples from the conditional distribution. We note that smaller values for \(\ell\) (shorter length-scales) imply more flexible GP models.}
    \label{fig:gp_prior}
\end{figure}

In order to understand the behaviour of \(\boldsymbol R\) different values for \(\ell\), we examine samples drawn from the prior distribution on \(\boldsymbol R\) conditioned on \(\sigma_f = 1\), \(R_0 = R_N = 1\), and \(N = 60\) for \(\ell \in \left\{ 10, 17.5, 25 \right\}\). Figure \ref{fig:gp_prior} illustrates the fact that a shorter length-scale results in a more flexible model for \(\boldsymbol R\), manifest as rapid transitions from high to low values of \(R_t\), and vice versa.

Specifically, when \(\ell = 10\) we observe that trends in \(R_t\) can change dramatically in only a few days, allowing for shifts in the reproduction number that are reversed within as little as ten days (see Figure \ref{fig:gp_prior}\textbf{a}). Such a model could accommodate an epidemic whereby cases surge twice within a two month period. Empirical data on COVID-19 in Ireland, where case numbers surged three times during the first 12 months of the COVID-19 epidemic, does not behave in this manner. Whether this is a consequence of human behaviour or some intrinsic property of the virus itself, it reflects the time-scale over which we expect to observe changes in the time-varying reproduction number. By this reasoning, \(\ell = 17.5\) is more consistent with empirical data (see Figure \ref{fig:gp_prior}\textbf{b}). Within this model, \(R_t\) remains stable over any given 2-3 week period, reflecting the time-scale on which non-pharmaceutical interventions are enforced. 
A longer length-scale, such as $\ell = 25$, sees similar conditions persist over the entire 60 days considered (see Figure \ref{fig:gp_prior}\textbf{c}). Such a model is unlikely to be flexible enough to fit \(R_t\) as epidemic conditions change. Thus, specifying a hyper-prior such that $\ell \sim \mathcal N \left( 17.5, 2.5 \right)$ seems a reasonable choice.

\section{Exploring changes in the generation interval pmf} \label{app:gen_int}

\begin{figure}[t]
\centering
\begin{subfigure}{\textwidth}
    \includegraphics[width=0.9\textwidth]{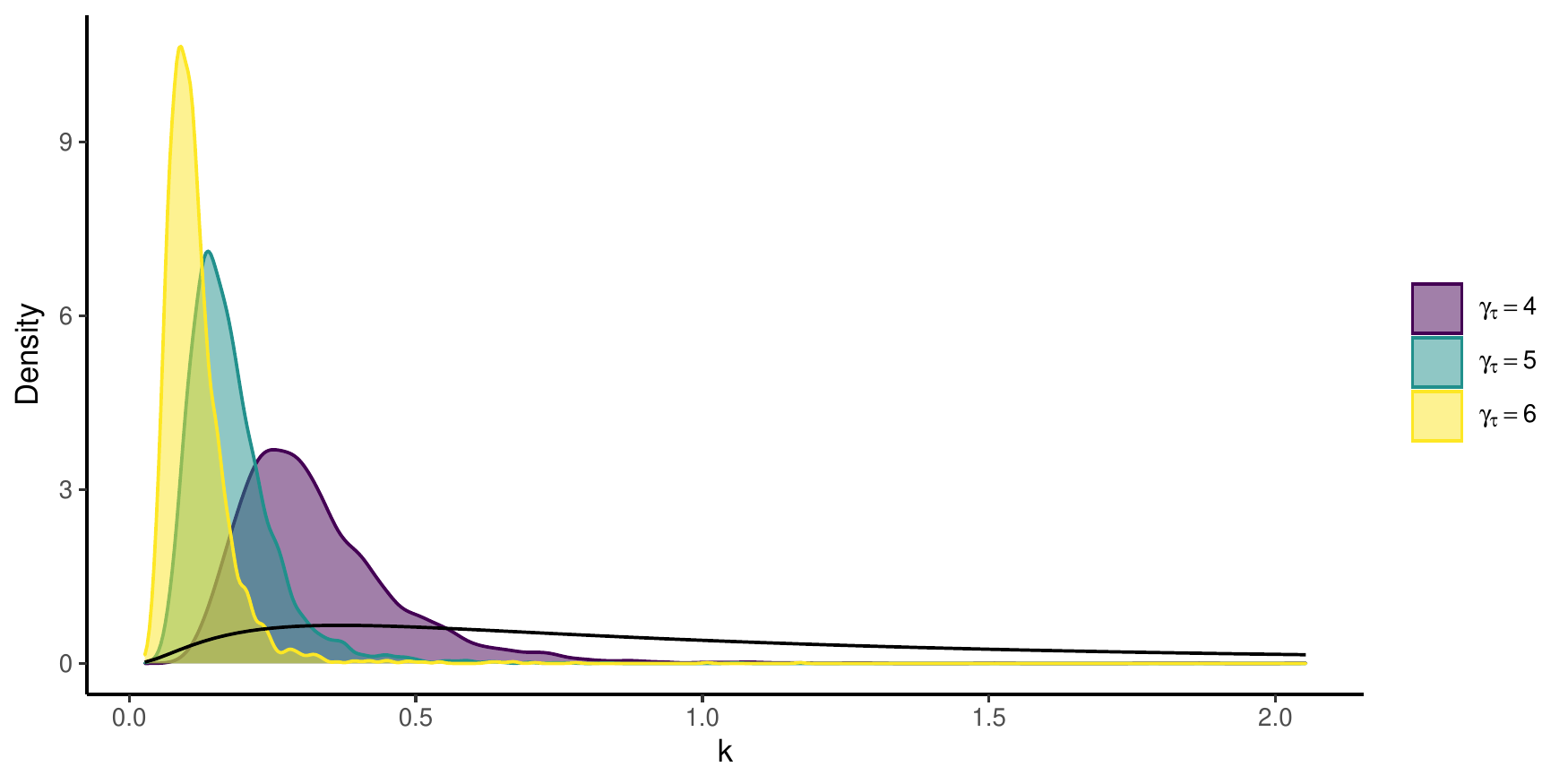}
    \caption{Inference for $k$}
    \label{fig:alt_k_figure}
\end{subfigure}

\begin{subfigure}{\textwidth}
    \includegraphics[width=0.9\textwidth]{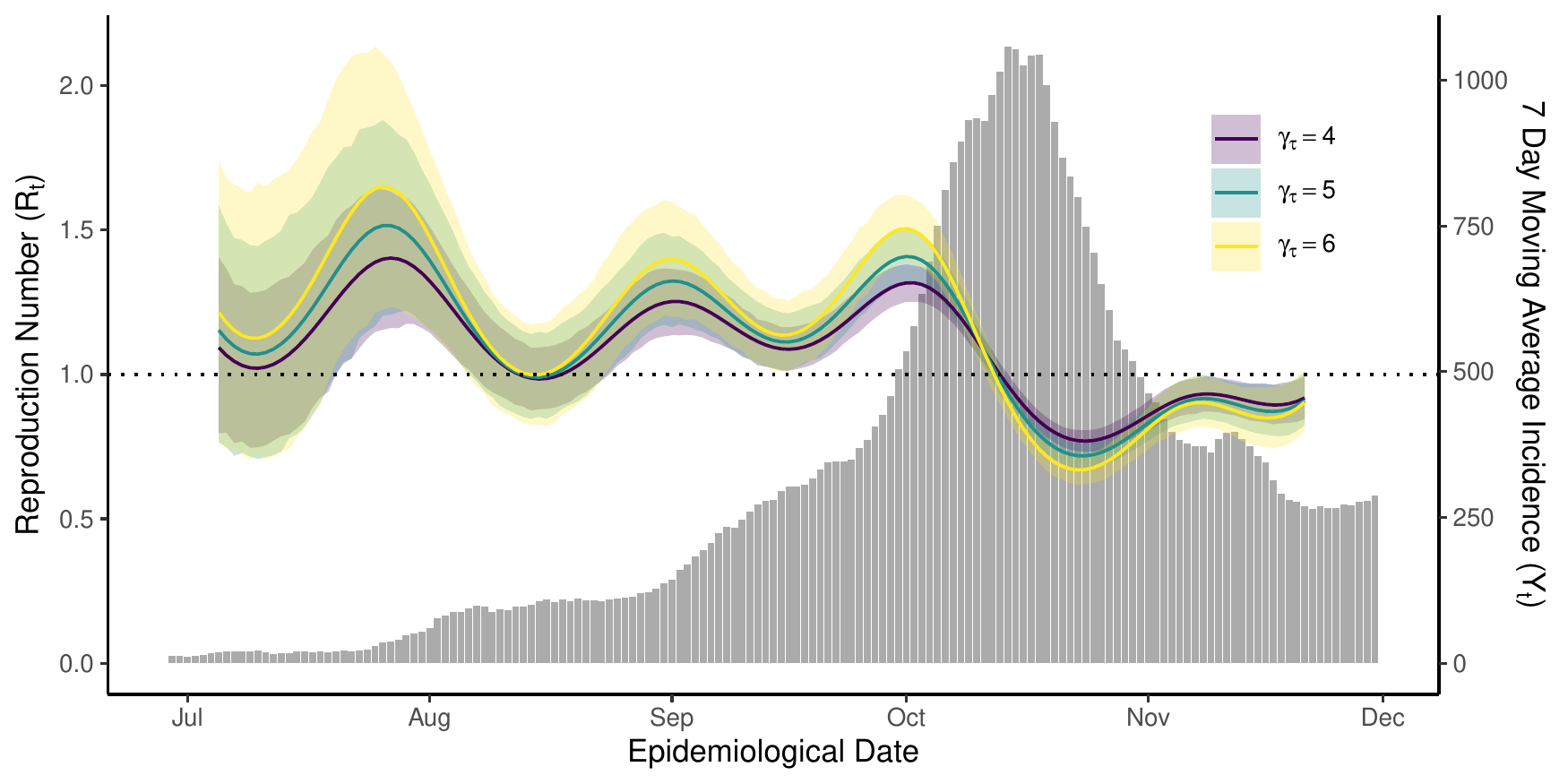}
    \caption{Inference for R}
    \label{fig:alt_R_figure}
\end{subfigure}
        
\caption{This analysis considers three candidate generation interval pmfs such that $\boldsymbol \omega$ defined in (\ref{eq:generation_interval_pmf}) is parameterised by $(\gamma_\tau, \sigma_\tau) \in \left\{(4, 2), (5, 2.5), (6, 3)\right\}$. Figure (a) presents the sampled posterior over $k$, while Figure (b) presents posterior inference for $\boldsymbol R$. We see that, while changes in $\omega$ do have an impact on inference, our overall conclusions on the heterogeneity of disease transmission remain broadly similar.}
\label{fig:alt_figure}
\end{figure}

In this Appendix, we consider how changes in the generation interval pmf $\boldsymbol \omega$ may effect our inference for $\boldsymbol R$ and $k$. We consider three candidate generation intervals chosen from the range for COVID-19 reported in the literature \citep{du2020serial, ganyani2020estimating, griffin2020rapid, rai2020estimates}, such that (\ref{eq:generation_interval_pmf}) is parameterised by $(\gamma_\tau, \sigma_\tau) \in \left\{(4, 2), (5, 2.5), (6, 3)\right\}$.
We sample from (\ref{eq:joint_conditional}) for each model in the manner described in the main text and verify our posterior samples satisfy diagnostic checks and tests for convergence.
The results of this analysis are presented in Table \ref{tab:alt_k} and Figure \ref{fig:alt_figure}.

\begin{table}[ht]
\centering
\begin{tabular}{|c|ccc|}
  \hline
  $(\gamma_\tau, \sigma_\tau)$ & (4, 5) & (5, 2.5) & (6, 3) \\
  \hline
  $k$ & 0.33 (0.15) & 0.18 (0.8) & 0.12 (0.06) \\ 
   \hline
\end{tabular}
\caption{The posterior mean (standard deviation) for $k$ under each candidate generation interval. We observe an inverse relationship between $k$ and $\gamma_\tau$.}
\label{tab:alt_k}
\end{table}

As described by \cite{gostic2020practical}, Figure \ref{fig:alt_R_figure} illustrates that larger mean generation intervals force estimates for $R_t$ away from 1; however, for each specification of $\boldsymbol \omega$, 95\% credible intervals overlap. Of more interest is the impact of changes in $\boldsymbol \omega$ on inference for $k$, presented in Figure \ref{fig:alt_k_figure} and Table \ref{tab:alt_k}. We observe that smaller values of $\gamma_\tau$ are associated with larger $k$. This analysis suggests that, for example, the 20\% most infectious individuals give rise to 65--85\%, 80--95\%, and 90-99\% of expected secondary infections when $\gamma_\tau$ is 4, 5, and 6 days, respectively. All three candidate generation intervals imply a high level of heterogeneity and lead us to broadly similar conclusions, although longer generation intervals are associated with more heterogeneous disease transmission. As such, we have presented only the model where $\gamma_\tau = 5$ in the main text, although future work may consider model averaging to allow for uncertainty in $\boldsymbol \omega$.
This analysis suggests that an inverse relationship exists between $k$ and $\gamma_\tau$, although a more detailed study lies beyond the scope of this report and is left for future work.

\end{document}